\documentclass[preprint,longtitle]{elsarticle}

\usepackage[english, brazil]{babel}
\usepackage[utf8]{inputenc}
\usepackage{amsmath}
\usepackage{float}
\usepackage{amssymb}
\usepackage{epsfig,psfrag,latexsym,subfigure}
\usepackage{graphics} 
\def\1{{1\kern-.3468em{ 1}}}
\usepackage[active]{srcltx}
\usepackage[color,final]{showkeys}

\biboptions{authoryear}
\usepackage{natbib}
\usepackage{tikz}
\usetikzlibrary{arrows,positioning}
\usepackage{wrapfig}

\newtheorem{theo}{Theorem}
\newenvironment{proof}[1][Proof]{\begin{trivlist}
\item[\hskip \labelsep {\bfseries #1}]}{\end{trivlist}}

\begin{document}
\selectlanguage{english}

\begin{frontmatter}

\title{\LARGE \bf A Novel Stochastic Epidemic Model with Application to COVID-19}
\author[Soton,COPPE]{Edilson F. Arruda}
\ead{e.f.arruda@southapton.ac.uk}
\author[COPPE]{Rodrigo e A. Alexandre}
\ead{alvim.rodrigo@yahoo.com.br}
\author[LNCC]{Marcelo D. Fragoso}
\ead{frag@lncc.br}
\author[Unicamp]{João B. R. do Val}
\ead{jbosco@fee.unicamp.br}
\author[Kerala]{Sinnu S. Thomas}
\ead{sinnu.thomas@iiitmk.ac.in}

\address[Soton]{Department of Decision Analytics and Risk, Southampton Business School, University of Southampton, 12 University Rd, Southampton SO17 1BJ, UK}
\address[COPPE]{Alberto Luiz Coimbra Institute-Graduate School and Research in Engineering, Federal University of Rio de Janeiro. CP 68507, Rio de Janeiro 21941-972, Brasil}
\address[LNCC]{National Laboratory for Scientific Computation, Av. Gett\'ulio Vargas 333, Quitandinha, Petrópolis RJ 25651-075, Brasil}
\cortext[cor1]{Corresponding author. Tel.: +44 023 8059 7677}
\address[Unicamp]{School of Electrical Engineering, University of Campinas, Av. Albert Einstein 400, Cidade Universitária, Campinas SP 13083-852, Brasil}
\address[Kerala]{School of Computer Science and Engineering, Kerala University of Digital Sciences, Innovation and Technology, Technocity, Mangalapuram
Thonnakkal PO Thiruvananthapuram, Kerala-695317, India}

\begin{abstract}
In this paper we propose a novel  SEIR stochastic epidemic model. A distinguishing feature of this new model is that it allows us to consider a set up under general latency and infectious period distributions. To some extent, queuing systems with infinitely many servers and a Markov chain {\em with time-varying transition rate} are the very technical underpinning of the paper.  Although more general, the Markov chain is as tractable as previous models for exponentially distributed latency and infection periods. It is also  significantly simpler and more tractable than semi-Markov models with a similar level of generality. Based on the notion of stochastic stability, we derive a sufficient condition for a shrinking epidemic in terms of the queuing system's occupation rate that drives the dynamics. Relying on this condition, we propose a class of ad-hoc stabilising mitigation strategies that seek to keep a balanced occupation rate after a prescribed mitigation-free period. We validate the approach in the light of recent data on the COVID-19 epidemic and assess the effect of different stabilising strategies. The results suggest that it is possible to curb the epidemic with various occupation rate levels, as long as the mitigation is not excessively procrastinated.
\end{abstract}
\end{frontmatter}


\section{Introduction \label{sec:intro}}

Classical epidemic models \citep[e.g.,][]{Ross1916,Kermack1927,Hethcote2000} are powerful tools to understand the spread of diseases and support public health policies. However, these models are deterministic and consequently do not capture the underlying uncertainties of the spread. Stochastic epidemic models \citep{Allen2008,Britton2010} were therefore designed to better capture some of these uncertainties and provide more realistic support for decision making.

Perhaps due to its simplicity, the \emph{susceptible, infected, removed} (SIR) is arguably the most utilised class of stochastic epidemic models. \citet{Trapman2009} made use of this framework to demonstrate the advantage of an M/G/1 queuing model to estimate the size of an epidemic at the time of detection. Some time later, classical M/M/S queues were utilised to estimate the whole outbreak of the Ebola virus \citep{Dike2016}. More recently, \citet{Barraza2020} employed pure Birth processes to fit data from the initial stages of the COVID-19 epidemic. Underpinning the analysis is the theory of Markov processes \citep{Bremaud1999}, used to model the transition of individuals among the SIR populations.

Markov models provide a powerful analytical framework for SIR models, allowing for example the treatment of non-homogeneous populations \citep{Lopez2016}. One of the limitations, however, is that the duration of the infectious period is assumed exponential, thus narrowing the technique's applicability \citep{Clancy2014,Corral2017}. A possible alternative is to develop more complex block-structured Markov chains that can mimic certain types of non-exponential infection times \citep{Lefevre2020,Islier2020}. Albeit limited, the added flexibility comes at the price of less interpretable and tractable models.

A thorough treatment of general infection times demands a class of semi-Markov processes known as piecewise-deterministic processes (PDP) \citep{Davis1993}. \citet{Clancy2014} uses such a class and martingale theory to derive the distribution of the number of infected individuals throughout the pandemic under generally distributed infection times. Later, \citet{Corral2017} utilised a similar model to obtain the distribution of the number of secondary cases due to an infected individual. However, it requires a memory of the disease progression of all currently infected individuals, which impacts the model's analytical and computational tractability and limits its use.


Inspired by the recent COVID-19 outbreak, this paper uses the SEIR (\emph{susceptible, exposed, infected, removed}) framework, which is analogous to SIR except for considering a latency period, whereby a recently infected individual does not manifest the disease nor can transmit it for a limited period after acquiring the illness. In addition to being more general, such a modelling choice is adequate for the COVID-19 epidemic, given its non-negligible latency period \citep{Backer2020}. Examples of stochastic SEIR models \citep{Artalejo2015,Lopez2017,Amador2018} assume exponentially distributed latency and infection periods. The first work concerns the duration of the outbreak, the second the transmission rate per infected person and the latter the epidemic size.

This paper builds upon previous literature \citep{Trapman2009} with innovative use of two $M_t/G/\infty$ models to describe the epidemic's stochastic
behaviour. However, whilst the $M/G/1$ queue in \citep{Trapman2009} models the epidemic up to the time of detection, the proposed model covers the whole outbreak. Such a generalisation hinges on two unique novelties. Firstly, we describe the input process as a time-varying Poisson process, which enables us to describe the variation of infection rates as a function of the system's dynamics. Secondly, by realising that there are no limits in the number of new infections and considering that conditions progress in parallel, we select a queuing model with infinitely many servers.

To the best of our knowledge, this is the first work to introduce a time-varying Markov chain capable of emulating a stochastic epidemic model with general latency and infection times. This innovation relies on the results of \citet{Eick1993}, that demonstrate that the output process of $M_t/G/\infty$ queues comprises a series of time indexed Poisson variables. Albeit as tractable as the models that assume exponential (memoryless) infection and latency periods \citep[e.g.,][]{Artalejo2015,Lopez2017,Amador2018}, the proposed framework is more general than the complex block-structured models \citep[e.g.,][]{Lefevre2020,Islier2020} as it does not impose any assumption on the infection and latency period distributions.

Even if our approach's level of generality is matched in part by previous PDP models within the more restricted SIR framework \citep{Clancy2014,Corral2017}, these require memory of the disease progression of all individuals in the \emph{infected} population. Of course, this is infeasible for all but tiny population sizes and limits such models' applicability to support decision making. In contrast, the proposed framework considers general latency and infection periods by keeping track solely of the new expositions within a complete cycle of the disease: from catching the virus to entering the removed population.

In addition to the methodological innovations, we propose a novel strategy to curb the epidemic that is based on the classical occupation rate of the $M_t/G/\infty$ exposition-to-removal queue. We claim that this measure is analogous to the deterministic reproduction number as it also provides a dynamic estimate of the epidemic's short-term growth. The strategy relies on mitigating actions specially tailored to maintain the occupation rate $\rho$ of the system as close as possible to a prescribed level $0 < \bar \rho < 1$ at all times and thus ensure a shrinking epidemic. We use COVID-19 data from the literature to validate the strategy by means of experiments designed to illustrate the utility of the approach whilst providing invaluable insights into the system's response to the proposed class of mitigation policies. The results demonstrate that by maintaining adequate occupation level targets, the epidemic can be conquered, as long as the mitigation is not excessively delayed.

This work is organised as follows. Section \ref{sec:form} introduces the mathematical formulation of the proposed SEIR model, starting with the proposed $M_t/G/\infty$ queues. We then use these queues' resulting input and output processes to propose a continuous time Markov chain to describe the epidemic's evolution. Section \ref{sec:stability} analyses stochastic stability and makes use of the results to propose a class of ad-hoc mitigation strategies to curb the epidemic. Section \ref{sec:numex} employs COVID-19 data from the literature to establish a set of experiments designed to illustrate the performance of the model in a real-world setting. The experiments also highlight the effectiveness of prescribed mitigation policies belonging to the class introduced in the previous Section. Finally, Section \ref{sec:conc} concludes the manuscript.

\subsection{A Few Notations} \label{notations} 

Throughout this paper we shall be using the following notation.  Let $\mathbb{Z_+}$ be the usual set of non-negative integers and $\mathcal{N}=\{0,1,2,3,...N\} \subset   \mathbb{Z_+}$, where $N$ is a finite number. Consider $\Omega=\mathcal{N}^4$ and define the probability space $(\Omega,\mathcal{F},\mathbb{P})$. In addition, $\mathcal{E}(\cdot)$ stands for the mathematical expectation.

\section{Mathematical formulation \label{sec:form} }

\begin{figure}[htb]
\centering \begin{tikzpicture}[node distance=1cm,auto,>=latex',every node/.append style={align=center},int/.style={draw, minimum size=1cm}]
    \node [int] (S)             {$S$};
    \node [int, right=of S] (E) {$E$};
    \node [int, right=of E] (I) {$I$};
    \node [int, right=of I] (R) {$R$};
    \coordinate[right=of I] (out);
    \path[->, auto=false] (S) edge node {$\beta I$ \\[.2em]} (E)
                          (E) edge node {$\sigma^{-1}$ \\[.2em]} (I)
                          (I) edge node {$\gamma^{-1}$ \\[.2em] } (out);
\end{tikzpicture}
\caption{The classical SEIR model \label{fig:seir}}
\end{figure}
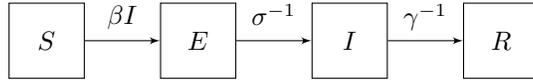

This section proposes a stochastic dynamic formulation in a probability space   $(\Omega,\mathcal{F},\mathbb{P})$. It relies on the classical deterministic SEIR epidemiological model \citep{Ross1916,Kermack1927,Hethcote2000}, thus depicted in Figure \ref{fig:seir} and in Table \ref{tab:parameters}.

The model comprises four compartments, namely: $(i)$ susceptible, $(ii)$ exposed,  $(iii)$ infected and $(iv)$ removed. \emph{Susceptible} individuals can be infected if they come in contact with infected individuals. The \emph{exposed} population, meanwhile, have been infected but the infection is still latent. In that phase, they are not able to spread the disease and do not manifest any symptom. Once the latency period is over, exposed persons become \emph{infected} and may present symptoms. Finally, infected individuals either die as a result of the disease or become immune to the disease. In both cases, they migrate to the \emph{removed} population, which consists of individuals that can no longer contract nor spread the disease.

In the deterministic model in Figure \ref{fig:seir}, susceptible individuals become ill at a rate $\beta >0$ upon making contact with infected (infectious) individuals, thus resulting in an infection rate $\delta = \beta S I$. The individuals who acquire the condition immediately become exposed and initiate their latency period, which lasts on average $\sigma$ units of time. Upon completing the latency period, exposed individuals become infectious for an average of $\gamma$ time units. After that, they enter the removed population either due to acquired immunity or death. Table \ref{tab:parameters} below presents the parameters of the deterministic model:
\begin{center}
\begin{table}[htb]
\caption{\centering Parameters of SEIR dynamics. \label{tab:parameters}}
\centering \scriptsize{
\begin{tabular}{|c|l|c|} \hline \hline
\textbf{Parameter} & \textbf{Description} & \textbf{Unit} \\ \hline
$\beta$ & Transmission rate & transmissions/encounter \\ \hline
$\sigma$ & Latency period & $\text{days}$ \\ \hline
$\gamma$ & Recovery period & $\text{days}$ \\ \hline
\end{tabular}
}
\end{table}
\end{center}

\subsection{Stochastic formulation \label{sec:stform} }

Although the model in Figure \ref{fig:seir} is invaluable to understand the underlying process, the dynamics are truly stochastic. Both the latency and the recovery period are, indeed, stochastic variables. See for example \citet{Backer2020} and \citet{Verity2020} for estimations of the latency and recovery periods at the early stages of the COVID-19 epidemic. In this Section, use the insights of \citet{Trapman2009} to model the SEIR dynamics as a queue with infinite service capacity.

\begin{figure}[!htb]
	\centering
\tikzset{every picture/.style={line width=0.75pt}} 
\begin{tikzpicture}[x=0.75pt,y=0.75pt,yscale=-1,xscale=1]

\draw  [fill={rgb, 255:red, 155; green, 155; blue, 155 }  ,fill opacity=1 ] (111,119.81) -- (194,119.81) -- (194,148) -- (111,148) -- cycle ;
\draw    (27,131.81) -- (105,132.97) ;
\draw [shift={(107,133)}, rotate = 180.85] [color={rgb, 255:red, 0; green, 0; blue, 0 }  ][line width=0.75]    (10.93,-3.29) .. controls (6.95,-1.4) and (3.31,-0.3) .. (0,0) .. controls (3.31,0.3) and (6.95,1.4) .. (10.93,3.29)   ;
\draw    (199,134) -- (251,134.96) ;
\draw [shift={(253,135)}, rotate = 181.06] [color={rgb, 255:red, 0; green, 0; blue, 0 }  ][line width=0.75]    (10.93,-3.29) .. controls (6.95,-1.4) and (3.31,-0.3) .. (0,0) .. controls (3.31,0.3) and (6.95,1.4) .. (10.93,3.29)   ;
\draw  [fill={rgb, 255:red, 155; green, 155; blue, 155 }  ,fill opacity=1 ] (258,121.81) -- (350,121.81) -- (350,150) -- (258,150) -- cycle ;
\draw    (355,135) -- (409,136.93) ;
\draw [shift={(411,137)}, rotate = 182.05] [color={rgb, 255:red, 0; green, 0; blue, 0 }  ][line width=0.75]    (10.93,-3.29) .. controls (6.95,-1.4) and (3.31,-0.3) .. (0,0) .. controls (3.31,0.3) and (6.95,1.4) .. (10.93,3.29)   ;
\draw  [color={rgb, 255:red, 126; green, 211; blue, 33 }  ,draw opacity=1 ] (54,111.16) .. controls (54,104.45) and (59.45,99) .. (66.16,99) -- (218.84,99) .. controls (225.55,99) and (231,104.45) .. (231,111.16) -- (231,147.65) .. controls (231,154.37) and (225.55,159.81) .. (218.84,159.81) -- (66.16,159.81) .. controls (59.45,159.81) and (54,154.37) .. (54,147.65) -- cycle ;
\draw  [color={rgb, 255:red, 74; green, 144; blue, 226 }  ,draw opacity=1 ] (46,100.81) .. controls (46,89.77) and (54.95,80.81) .. (66,80.81) -- (371,80.81) .. controls (382.05,80.81) and (391,89.77) .. (391,100.81) -- (391,160.81) .. controls (391,171.86) and (382.05,180.81) .. (371,180.81) -- (66,180.81) .. controls (54.95,180.81) and (46,171.86) .. (46,160.81) -- cycle ;

\draw (60,110) node [anchor=north west][inner sep=0.75pt]  [color={rgb, 255:red, 139; green, 87; blue, 42 }  ,opacity=1 ]  {$\lambda ( t)$};
\draw (198,112) node [anchor=north west][inner sep=0.75pt]  [color={rgb, 255:red, 139; green, 87; blue, 42 }  ,opacity=1 ]  {$\delta _{e}( t)$};
\draw (146,125) node [anchor=north west][inner sep=0.75pt]   [align=left] {{\fontfamily{helvet}\selectfont {\large \textbf{E}}}};
\draw (302,127) node [anchor=north west][inner sep=0.75pt]   [align=left] {\textbf{{\large {\fontfamily{helvet}\selectfont I}}}};
\draw (353,114) node [anchor=north west][inner sep=0.75pt]  [color={rgb, 255:red, 139; green, 87; blue, 42 }  ,opacity=1 ]  {$\delta _{i}( t)$};
\draw (413,126) node [anchor=north west][inner sep=0.75pt]   [align=left] {{\large {\fontfamily{helvet}\selectfont \textbf{R}}}};
\draw (105,158) node [anchor=north west][inner sep=0.75pt]  [color={rgb, 255:red, 126; green, 211; blue, 33 }  ,opacity=1 ] [align=left] {Queue 2};
\draw (183,186) node [anchor=north west][inner sep=0.75pt]   [align=left] {\textbf{{\large \textcolor[rgb]{0.29,0.56,0.89}{Queue 1}}}};
\draw (114,99) node [anchor=north west][inner sep=0.75pt]  [color={rgb, 255:red, 139; green, 87; blue, 42 }  ,opacity=1 ]  {latency\ $( \sigma )$};
\draw (261,102) node [anchor=north west][inner sep=0.75pt]  [color={rgb, 255:red, 139; green, 87; blue, 42 }  ,opacity=1 ]  {recovery\ $( \gamma )$};
\end{tikzpicture}
\caption{\textbf{Queue 1:} $M_t/G/\infty$ Queue from Exposed to Removed. \textbf{Queue 2:} $M_t/G/\infty$ Queue from Exposed to Infected}
	\label{fig:total}
\end{figure}
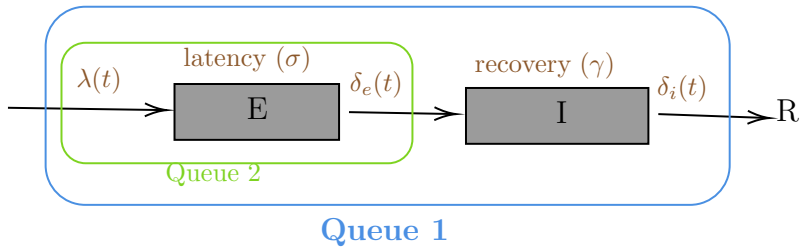

\textbf{Queue 1} in Figure \ref{fig:total} represents an individual's trajectory from acquiring the disease and therefore becoming exposed, to their removal of the system. Because the individuals' trajectories are assumed independent, the system can be modelled as a $M_t/G/\infty$ queue \citep{Eick1993}, since there is no upper limit on the number of simultaneous infections. The service time is the sum of two generally distributed random variables $\sigma$ and $\gamma$, representing the latency and recovery periods, see Table \ref{tab:parameters}; the latter corresponds to the length of the infectious period. 


Following a previous study of the outset of epidemic \citep{Trapman2009}, we model the arrival rate as a Poisson random variable with a rate proportional to the maximum number of encounters between healthy and infected individuals. However, the proposed model innovates by considering a time-varying input rate $\lambda(t) = \beta S(t) I(t)$ to cover the epidemic's evolution  over time. Observe in Figure \ref{fig:total} that each departure from Queue 1 represents a decrease of one infected individual. In turn,  each arrival corresponds to a new exposition, i.e. a susceptible individual contracting the disease in its latent stage. The compartments $E$, $I$, and $R$ denote the corresponding SEIR populations. 

Also in Figure \ref{fig:total}, \textbf{Queue 2} is a second $M_t/G/\infty$ queue that covers the \emph{latency} period $\sigma$ only. Each departure of this queue increases the infected population by one.
%

The queues in Figure \ref{fig:total} are drawn from \citet{Eick1993}. However, the proposed epidemic model is more general since the input rate $\lambda(t)$ is not a deterministic function of time. Instead, it depends on the number of infected and susceptible individuals at time $t \ge 0$. As previously stated, stochastic jumps in the number of infected individuals depend upon the departure processes of the two queues introduced in Figure \ref{fig:total}. On the other hand, the susceptible population is non-increasing over time and decreases with each new arrival.

\subsubsection{The arrival and departure processes}

We start by modelling the arrival process shared by both queues. At any given time $t \ge 0$, the rate of contagion (exposition) is given by
\begin{equation} \label{eq:inputrate}
    \lambda(t) = \beta S (t) I(t), \; t \ge 0,
\end{equation}
where $\beta$ is a scalar parameter, $S(t)$ and $I(t)$ are the sizes of the susceptible and infected populations at time $t$, respectively.

By design, the output of \textbf{Queue 2} - in Figure \ref{fig:total} - is the number of new infections, i.e. individuals become \emph{infected} when they depart this queue.  Let $\delta_e(t)$ be a random variable denoting the number of departures from Queue 2 at time $t \ge 0$. According to \citet{Eick1993}, $\delta_e(t)$ is a Poisson random variable with mean:
%
%
\begin{equation} \label{eq:inputinfected}
\overline{\delta_e}(t) =  \mathcal{E}(\lambda  \,( t - \sigma) \,).
\end{equation}
This is because the latency period $\sigma$ is also the service time of Queue 2.

Let $\delta_i(t)$ be a random variable representing the number of departures from \textbf{Queue 1} at time $t \ge 0$. Each departure is a newly removed individual and the total time from exposition to removal is the sum of random variables $\sigma$ and $\gamma$, which is also the service time of Queue 1. Once again, the results of \citet{Eick1993} imply that $\delta_i(t)$ is a Poisson random variable with mean:
\begin{equation} \label{eq:outputinfected}
\overline{ \delta_i}(t)  = \mathcal E(\lambda( \,t - ( \sigma + \gamma ) \,) ).
\end{equation}

\subsubsection{Markov formulation with time-varying transition rate\label{sec:tvMarkov}}

We make use of the arrival and departure rates of Queues 1 and 2 in Eq. \eqref{eq:inputrate}-\eqref{eq:outputinfected} to define a time-varying Markov process $X_t, \, t \ge 0$  that describes the evolution of the populations in the SEIR compartments. At any time $t \ge 0$, $X(t) = (\, S(t), \, E(t), \, I(t), R(t) \, ) \in \Omega$ is the state of process $X_t, \, t \ge 0$. By definition, $X_t, \, t \ge 0$ will be subject to random jumps that happen whenever either of these events occur: $i)$ a new exposition, $ii)$ a new departure from Queue 2 or $iii)$ a new departure from Queue 1.

To describe the dynamics of  the process $X_t, \, t \ge 0$, we make use of the fact that the input and output of both queues are described by Poisson variables \citep{Eick1993} at any given time $t \ge 0$. Consequently, the time until the next event will be exponentially distributed and the total jump rate  at time $t \ge 0$ is:
\begin{equation} \label{eq:ratelamb}
    \Lambda(t) = \lambda(t) + \overline{\delta_e}(t) + \overline{\delta_i}(t).
\end{equation}
Let $\{\tau_0, \, \tau_1, \ldots\}$ be the sequence of jumps in the system, with $\tau_0 \equiv 0$ and $\tau_{k+1} > \tau_k, \, \forall k \ge 0$. Now, assume a jump occurs at time $t = \tau_k$. Then, the following holds:

\begin{multline} \label{eq:transitons}
P(\, X(t^+) = Y | X(t) = (S(t), \, E(t),\, I(t), \, R(t)) ), \,t=\tau_k  \,) \\
=\begin{cases} 
\dfrac{\lambda(t)}{\Lambda(t)} & \text{if} \; Y = (S(t) - 1, \; E(t)+1,\; I(t), \; R(t)) \, ) \\
\dfrac{\overline{\delta_e}(t)}{\Lambda(t)} & \text{if} \; Y = (S(t) , \; E(t)-1,\; I(t) + 1, \; R(t)) \, ) \\
\dfrac{\overline{\delta_i}(t)}{\Lambda(t)} & \text{if} \; Y = (S(t) , \; E(t),\; I(t) - 1, \; R(t) + 1) \, ) \\
0 & \text{otherwise}.
\end{cases}
\end{multline}


The first expression on the right-hand side of Eq. \eqref{eq:transitons} corresponds to a new contagion/exposition, which happens at rate $\lambda(t)$, see Eq. \eqref{eq:inputrate}, and implies the transference of an individual from the susceptible to the exposed population. The second expression corresponds to a new infection that happens upon the departure of an individual from Queue 2, which occurs at rate $\overline{\delta_e}(t)$, see Eq. \eqref{eq:inputinfected}. In that case, this individual moves from the exposed to the infected population. Finally, the third possibility is the departure of an individual from Queue 1, which happens at rate $\overline{\delta_i}(t)$, see Eq. \eqref{eq:outputinfected}. In that case, this individual migrates from the infected to the removed population.

%
Finally, after the jump at $t = \tau_k, \, k \ge 0$, the value of the exposition rate $\lambda(t)$ also changes, and becomes:
\begin{equation}
    \lambda(t^+) = \beta S(t^+) I(t^+),
\end{equation}
where the new values on the right-hand side of the expression above vary as a function of the transition probabilities in \eqref{eq:transitons}. Clearly, $\lambda(t)$ remains unaltered between successive jumps. Given an initial distribution and the transition probability in \eqref{eq:transitons}, the process $X_t, \, t \ge 0$ is a continuous-time Markov chain.

\section{Stochastic stability and reproduction number \label{sec:stability}}

Whereas deterministic models rely on the evaluation of trivial equilibrium to derive stability conditions and the so-called reproduction number $R_0$ \citep[e.g.][]{Driessche2002}, the notion of stochastic stability \cite[e.g.,][]{meyn93} provides an ideal framework to evaluate the conditions for a receding epidemic as time elapses. Queuing theory connects this notion with the so-called occupation rate, a measure of the input to output ratio as time elapses \citep{Shortle2018}. 

Consider Queue 1 in Figure \ref{fig:total}. Assuming a very large population, the stability of such a queue hinges on the occupation rate \citep[e.g.,][]{Shortle2018}:
\begin{equation} \label{eq:rho}
    \rho(t) = \frac{\lambda(t)}{\overline{\delta_i}(t)},
\end{equation}
and can be ascertained if a finite time $\bar{t} \ge 0$ exists such that $\rho(t) < 1, \forall t \ge \bar t$. This guarantees that the system stabilises and the number of customers in the queue remains finite. With a finite population, however, stability is guaranteed because the number of infected individuals will remain within finite, albeit possibly large, bounds. In that case, we are interested in the trend of Queue 1, which indicates whether the epidemic is increasing or receding. Theorem \ref{theo:stability} below establishes the condition for a receding epidemic.

\begin{theo} \label{theo:stability}
Consider the Markov process described in Section  \ref{sec:tvMarkov} and assume that $\rho(t) < 1$ for all $t > \bar t \ge 0$, with $\bar t < \infty$. Then, it follows that:
\begin{equation} \label{eq:main}
     \mathcal E \left(\, E(t^+) + I(t^+)~ | ~X(t)~ ; ~  t={\tau}_{k}  \, \right)  < E(t) + I(t), \; \forall t = \tau_k > \bar t, \, k \ge 0,
\end{equation}
where $\{\tau_0, \, \tau_1, \ldots\}$ is the sequence of jumps in the system, as defined in Section \ref{sec:tvMarkov}.

\end{theo}
\begin{proof}
%
From \eqref{eq:transitons} we have that:
\small{
\begin{align*}
    \mathcal E \left(\, E(t^+) + I(t^+)~ | ~X(t)~ ; ~  t={\tau}_{k}  \, \right) \; = & \frac{\lambda(t)}{\Lambda(t)} \left( E(t) + I(t) + 1 \right) + \\
   &  \frac{\overline{\delta_e}(t)}{\Lambda(t)} \left( E(t) + I(t) \right) + \frac{\overline{\delta_i}(t)}{\Lambda(t)} \left( E(t) + I(t) - 1 \right) \\
   = & E(t) + I(t)  + \frac{\lambda(t) - \overline{\delta_i}(t)}{\Lambda(t)},
\end{align*}
}
for all $t = \tau_k$. The last equality holds because $\Lambda(t) = \lambda(t) + \overline{\delta_e}(t) + \overline{\delta_i}(t)$ - Eq. \eqref{eq:ratelamb}. Now, Eq. \eqref{eq:main} follows by assuming $\rho(t) < 1$ for all $t > \bar t \ge 0$.
\end{proof}

As a consequence of Theorem \ref{theo:stability}, we can interpret $\rho(t)$ as the reproduction number of the system at time $t$ and $\rho(t) < 1 \, \forall t > \bar t \ge 0$ as a sufficient condition for the epidemic to stabilise and shrink. It is worth pointing out that the theorem indicates that the epidemic is receding only when the sum of exposed and infected individuals, i.e. all latent and manifested infections, decreases. It highlights that, although important, the number of infected individuals is not sufficient to establish the tendency of the epidemic. Indeed, it is quite intuitive that the latter depends upon the number of latent infections, which are bound to occur shortly and therefore cannot be ignored.

The model indicates that we have to consider the disease's whole cycle, from exposition to removal, to evaluate the epidemic trend, such as in Eq. \eqref{eq:rho}. It underscores the importance of keeping an accurate track of the epidemic's evolution through an efficient testing strategy.  The accurate evaluation of such a cycle's length will also be essential to evaluate the lags between mitigating actions and decreases in the number of infections and expositions. Although mitigation can prevent future expositions and infections, it has no effect on recent transmissions still in the latent stage, which will continue to manifest and may actually drive infection up in the first stages of the mitigation. This will become more evident in the experiments carried out in Section \ref{sec:numex}.



\subsection{Mitigation strategies \label{sec:mitigation}}

Theorem \ref{theo:stability} provides a basis for developing mitigation strategies to stabilise the epidemic, whereas the Markov model in Section \ref{sec:tvMarkov} enables us to evaluate the long-term effects of such strategies. Following the literature, we introduce the control (mitigating action) in the form of non-pharmaceutical interventions to limit the spread of the disease \citep[e.g.,][]{ferguson2020,Kantner2020,Tarrataca2021}. A control level $0 \le u(t) \le 1$ attains a reduction of $100 u(t)$ percentage points in the transmission rate at time $t \ge 0$, thus resulting in a controlled exposition rate:
\begin{equation} \label{eq:controlled_inputrate}
    \lambda(t, u(t)) = \beta (1-u(t)) S (t) I(t), \; t \ge 0.
\end{equation}

Let $\pi = \{u(t), \, t \ge 0\}$ be a mitigation strategy and let $\Pi$ denote the set of all feasible strategies $\Pi = \{\pi: 0 \le u(t) \le 1, \, \forall t \ge 0\}$. The dynamics of the system under any strategy $\pi \in \Pi$ can be evaluated by running the Markov chain $X_t, \, t \ge 0$ with the same underlying dynamics, but making $\lambda(t) = \lambda(t, u(t)), \, t \ge 0$.

We are particularly interested in the class of stabilising policies $\Pi^S \in \Pi$, such that:
\begin{equation} \label{eq:stabilising}
    u(t) = \begin{cases}
    0, & \text{if} \; t \le \bar t, \\
    \min \left[ \left( 1 - \displaystyle \frac{\bar \rho}{\rho(t)} \right), \;  0 \right], & \text{if} \; t > \bar t,
    \end{cases}
\end{equation}
where $\rho(t)$ is the uncontrolled occupation rate in Eq. \eqref{eq:rho}. It is easy to see that such policies lead to a controlled input rate $\rho_c(t) = \frac{\lambda(t, u(t))}{\overline{\delta_i}(t)} \le \bar \rho, \, \forall t > \bar t$, see Eq. \eqref{eq:controlled_inputrate}. Henceforth, $\bar \rho$ will be called \emph{target occupation level} and $\pi = (\bar t, \bar \rho)$ will describe any stabilising policy $\pi \in \Pi^S$ that satisfies Eq. \eqref{eq:stabilising}. In the next Section we will evaluate these policies in the light of the COVID-19 epidemic and explore the stabilising effect of parameters $\bar \rho$ and $\bar t$ in Eq. \eqref{eq:stabilising}. Observe that, because the population is finite, the effect of the control is highly sensitive to and can be limited by a delayed start of mitigation.

Observe, however, that the Markov model in Section \ref{sec:tvMarkov} is not limited to the proposed class of mitigation strategies. In fact, one can substitute any arbitrary mitigation policy $\pi \in \Pi$ in Eq. \eqref{eq:controlled_inputrate} and run the controlled Markov chain to evaluate the effect of such policy. To demonstrate that, our experiments will also include the \emph{on-off} lock-down policies proposed by \citet{Tarrataca2021}. Designed to control hospital bed's occupation, these policies trigger a full scale lock-down when infections surpass a prescribed upper bound; conversely, mitigating actions are lifted when infections fall bellow a prescribed lower bound.

\section{Numerical Experiments \label{sec:numex}}

This Section validates the proposed stochastic SEIR dynamic model in Section \ref{sec:tvMarkov} in the light of data from the COVID-19 epidemic. We replicate the parameters of \citet{Tarrataca2021} concerning the spread of the epidemic in Brazil. The parameters are listed in Table \ref{tab:paramsim}. 

\begin{center}
\begin{table}[htb]
\caption{\centering Parameters of the Simulation. \label{tab:paramsim}}
\centering \scriptsize{
\begin{tabular}{|c|l|c|} \hline \hline
\textbf{Parameter} & \textbf{Description} & \textbf{Value} \\ \hline
$P$ & Total population & $217\cdot10^6$ \\ \hline
$I(0)$ & Initial number of infected people & $2$ \\ \hline
$E(0)$ & Initial number of exposed people & $252$ \\ \hline
$S(0)$ & Initial number of susceptible people & $217\cdot10^6-254$ \\ \hline
$\beta$ & Transmission rate & $2.41\cdot10^{-9}$ \\ \hline
\end{tabular}
}
\end{table}
\end{center}

To complete the necessary data to run the model, we also need the distributions of the latency period $\sigma$ and of the latency plus infection period ($\sigma + \gamma)$, which represents the full disease cycle. The distributions are compatible with previous observations \citep{Backer2020,Verity2020} and assumed discrete to cope with the typical daily collection of data. Tables \ref{tab:Ltdist} and \ref{tab:Rmdist} unveil the distributions of $\sigma$ and $(\sigma + \gamma)$, respectively.

\begin{center}
\begin{table}[h!]
\caption{\centering Latency Period Distribution ($\sigma$). \label{tab:Ltdist}}
\centering \scriptsize{
\begin{tabular}{|c|c|c|c|c|c|c|c|c|c|c|c|c|c|c|c|c|} \hline \hline
\textbf{Day} & 0 & 1 & 2 & 3 & 4 & 5 & 6 & 7 \\ \hline
\textbf{Probability} & - & 0.0009 & 0.0056 & 0.0222 & 0.0611 & 0.1222 & 0.1833 & 0.2095 \\ \hline \hline
\textbf{Day} & 8 & 9 & 10 & 11 & 12 & 13 & 14 & \\ \hline
\textbf{Probability} & 0.1833 & 0.1222 & 0.0611 & 0.0222 & 0.0056 & 0.0009 & $6\cdot10^{-5}$ & \\ \hline
\end{tabular}
}
\end{table}
\end{center}

\begin{center}
\begin{table}[h!]
\caption{\centering Latency and Infection Period Distribution ($\sigma + \gamma$). \label{tab:Rmdist}}
\centering \tiny{
\begin{tabular}{|c|c|c|c|c|c|c|c|c|} \hline \hline
\textbf{Day} & 0 & 1 & 2 & 3 & 4 & 5 & 6 & 7\\ \hline
\textbf{Prob.} & - & - & - & - & - & - & - & - \\ \hline \hline
\textbf{Day} & 8 & 9 & 10 & 11 & 12 & 13 & 14 & 15\\ \hline
\textbf{Prob.} & - & - & $3.87\cdot10^{-7}$ & $4.84\cdot10^{-6}$ & $3.87\cdot10^{-5}$ & 0.0002 & 0.0010 & 0.0034 \\ \hline \hline
\textbf{Day}  & 16 & 17 & 18 & 19 & 20 & 21 & 22 & 23 \\ \hline
\textbf{Prob.}  & 0.0098 & 0.0233 & 0.0466 & 0.0792 & 0.1151 & 0.1439 & 0.1550 & 0.1439 \\ \hline \hline
\textbf{Day} & 24 & 25 & 26 & 27 & 28 & 29 & 30 & 31\\ \hline
\textbf{Prob.} & 0.1151 & 0.0792 & 0.0466 & 0.0233 & 0.0098 & 0.0034 & 0.0010 & 0.0002 \\ \hline \hline
\textbf{Day} & 32 & 33 & 34 & 35 &  &  &  & \\ \hline
\textbf{Prob.} & $3.87\cdot10^{-5}$ & $4.84\cdot10^{-6}$ & $3.87\cdot10^{-7}$ & $1.49\cdot10^{-8}$ &  &  &  &  \\ \hline
\end{tabular}
}
\end{table}
\end{center}

\subsection{The effect of delayed mitigation}

The first series of experiments examine policies $\pi = (\bar t, \bar \rho) \in \Pi^S$ that follow Eq. \eqref{eq:stabilising} with a fixed target occupation level $\bar \rho = 0.95$. The objective is to evaluate the effect of delaying the start of mitigating actions by varying the first parameter. Cases A-G in Table \ref{tab:simcases} feature different triggering times for the mitigating actions, starting with the case where no mitigation is enforced ($\bar t = \infty$). For each experiment, we ran a full realisation of the system for two years, starting from the outset of the epidemic.

\begin{center}
\begin{table}[hb]
\caption{\centering Simulated cases. \label{tab:simcases}}
\centering \scriptsize{
\begin{tabular}{|c|l|c|l|c|l|c|l|c|l|c|l|c|l|c|} \hline \hline
\textbf{Case} & A & B & C & D & E & F & G \\ \hline
\textbf{$\bar t$} & $\infty$ &  63 & 91 & 98 & 105 & 112 & 126 \\ \hline
\textbf{$\bar \rho$} & 0.95 & 0.95 & 0.95 & 0.95 & 0.95 & 0.95 & 0.95 \\ \hline
\end{tabular}
}
\end{table}
\end{center}

\begin{figure}[!htb]
	\centering
	\centerline{\includegraphics[width=\linewidth] {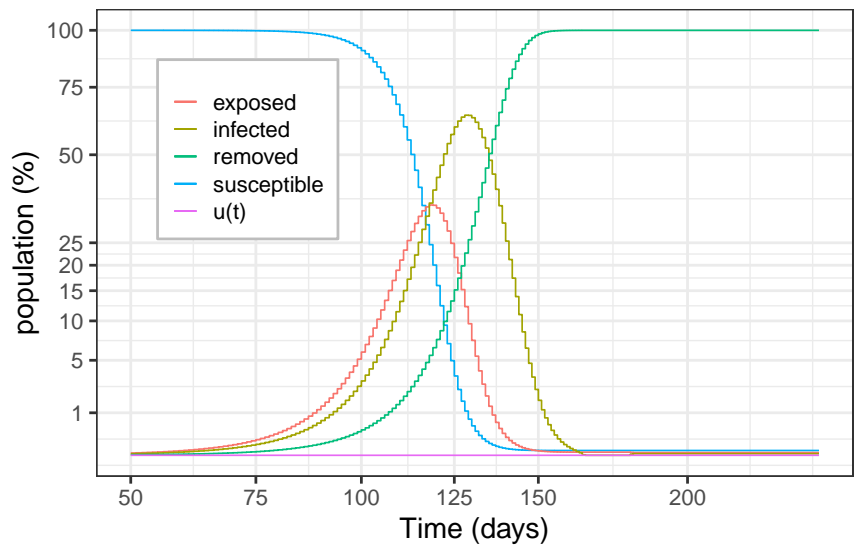}}
	\caption{\textbf{Epidemic evolution for Case A: the unmitigated spread}}
	\label{fig:case_a}
\end{figure}

\begin{figure}[!h]
	\centering
	\includegraphics[width=\linewidth] {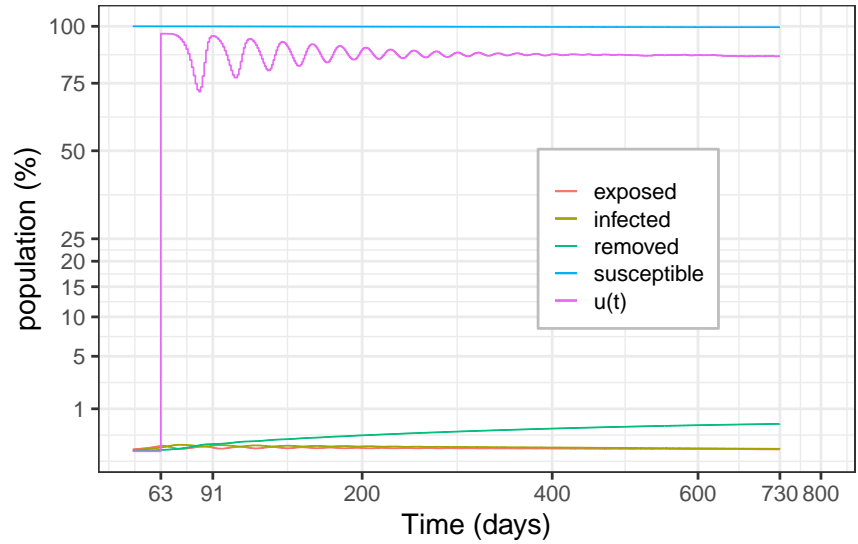}
	\includegraphics[width=\linewidth] {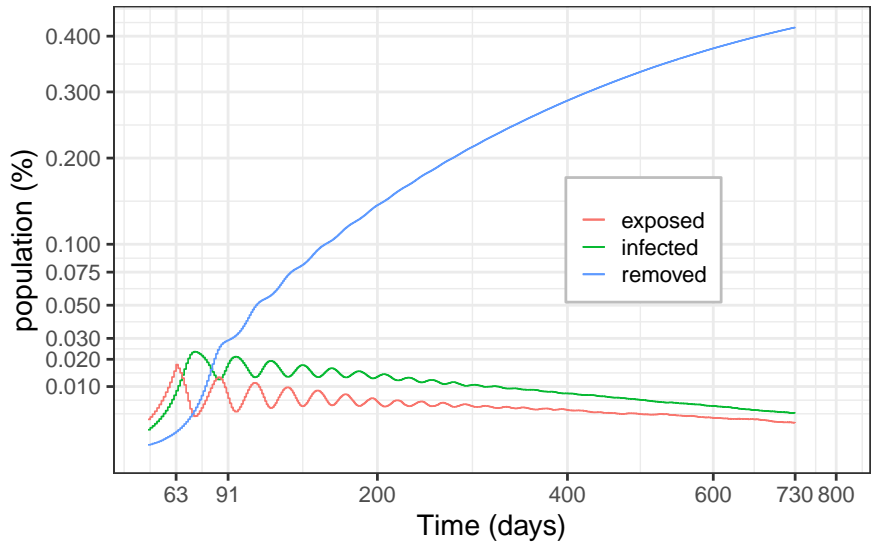}
	\caption{\textbf{Epidemic evolution for Case B}}
	\label{fig:case_b}
\end{figure}


Figure \ref{fig:case_a} shows the daily evolution of the epidemic for Case A, i.e. in the absence of mitigating actions. To facilitate the interpretation, the number of individuals in each SEIR compartment is depicted as a percentage of the total population. One can see that the epidemic spreads rapidly, with the number of infected individuals peaking around 64\% of the population after 128 days. The epidemic then starts to shrink due to the decrease in the susceptible population, until it finally subsides around the $165^{th}$ day.

Observe from Figure \ref{fig:case_b} that Case B starts the mitigation early in the epidemic, with low levels of infection. The upper plot conveys full results and in the lower plot we zoom in to detail the evolution of the exposed, infected and removed populations. The results illustrate the effectiveness of the mitigation policy. Observe that the mitigation policy successfully prevents the spread of the disease, maintaining 99.6\% of the population healthy over the whole two-year interval, whereas the removed population reaches 0.4\% at the end of that interval. Controlling the epidemic, however, demands very high levels of mitigation that oscillate and stabilise around $0.875$. It is noteworthy that deterministic SEIR-based optimal control approaches also required high levels of control to curb the COVID-19 epidemic, as can be seen in the relevant work of \citet{Perkins2020}.

\begin{figure}[!htb]
	\centering
	\centerline{\includegraphics[width=\linewidth] {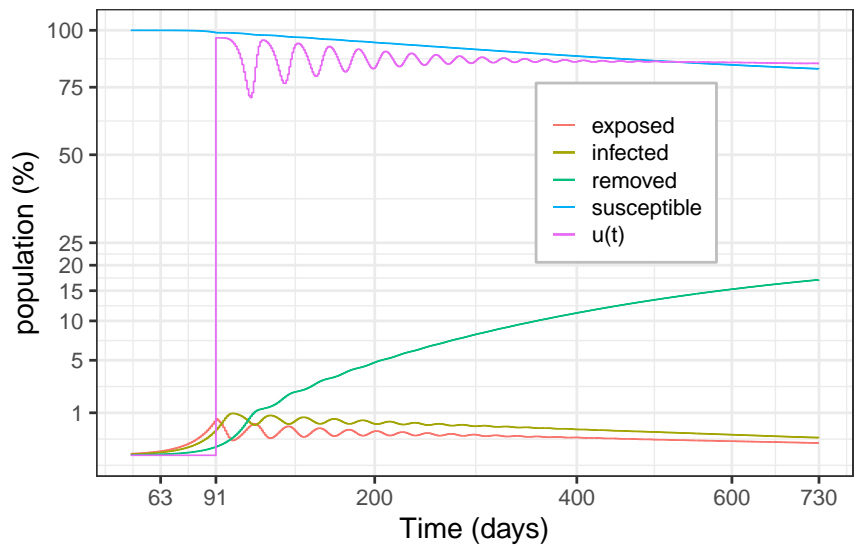}}
	\caption{\textbf{Epidemic evolution for Case C}}
	\label{fig:case_c}
\end{figure}

In Case C, the mitigation strategy is delayed for an extra three weeks with respect to Case B. One can see that the mitigating actions are similar, tend to stabilise around the same value and are equally sufficient to curb the epidemic. The difference with respect to Case B is a rather steep increase in the final number of individuals that catch the disease at some point, represented by the removed population. This number increases from $0.4\%$ in Case B to around 17\% in Case C, thus highlighting the sensitivity of the spread with respect to mitigation delay. The results also show that a three-week delay produces a large increment in the peak of infections, for even though this peak only reaches around 1\% of the population, it is about four times as large as it would be if the mitigation strategy started 21 days earlier.


\begin{figure}[!htb]
	\centering
	\centerline{\includegraphics[width=\linewidth] {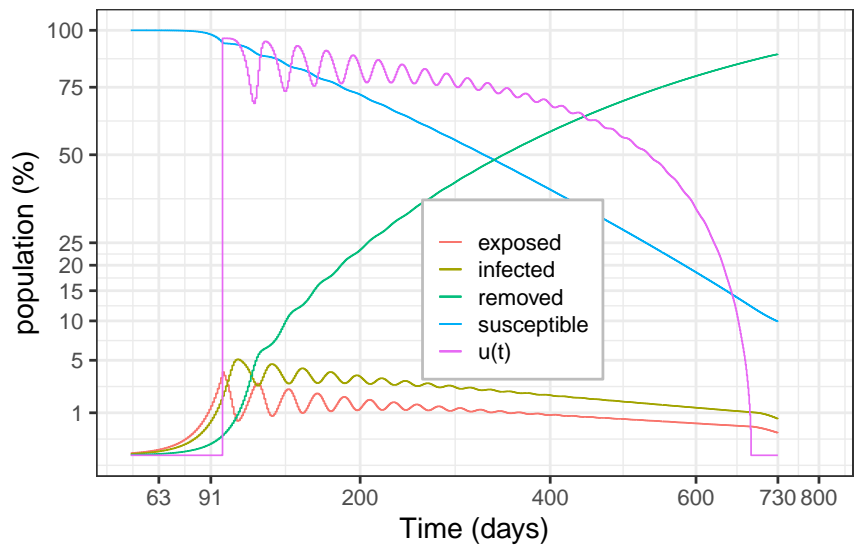}}
	\caption{\textbf{Epidemic evolution for Case D}}
	\label{fig:case_d}
\end{figure}

Figures \ref{fig:case_d} features the results for Case D. Despite an extra delay of one week with respect to Case C, it is still possible to flatten the curve and prevent an uncontrollable increase of the epidemic. Nonetheless, the extra week produces a peak of infections five times as large as in Case C, with 5\% of people being infected 107 days after the outset of the epidemic. The total removed population skyrockets from about 17\% in Case C to about 90\% in Case D, thus reinforcing the importance of early mitigation to prevent the spread of the disease and the exponential effect of extra delays in the system. Observe that, due to the spread of the disease and the corresponding decrease in the number of susceptible individuals, we are able to slowly relax the mitigation until the disease has contaminated about 50\% of the population. As the disease spreads further, we are able to rapidly relax the mitigation whilst still maintaining a steady decrease in the number of infected individuals. The mitigation is completely lifted just before the second anniversary of the outset of the epidemic.

\begin{figure}[!htb]
	\centering
	\centerline{\includegraphics[width=\linewidth] {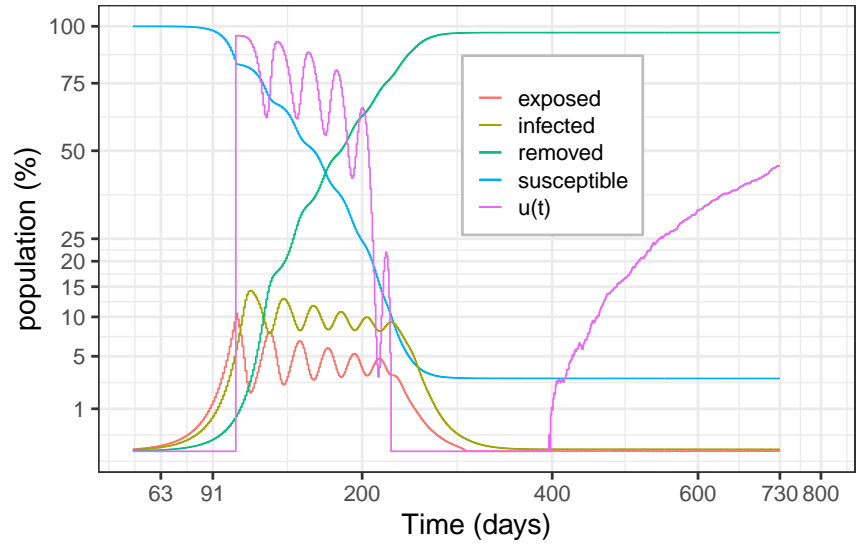}}
	\caption{\textbf{Epidemic evolution for Case E}}
	\label{fig:case_e}
\end{figure}

Figure \ref{fig:case_e} shows the results for Case E that provide further evidence of the deleterious effect of mitigation delay. A further delay of one week with respect to Case D results in approximately three times as many infected individuals at the peak, that occurs on day 114 and amounts to around 14\% of infected individuals. Due to the larger peak, we are able to relax the mitigation earlier and more rapidly, reaching a full relaxation just after seven months. However, since there are still susceptible individuals in the population, mitigation has to re-start from day 396 onwards to avoid a recrudescence of the epidemic. The final number of removed individuals indicates that the epidemic affects about 97.5\% of the population within two years.

\begin{figure}[!htb]
	\centering
	\centerline{\includegraphics[width=\linewidth] {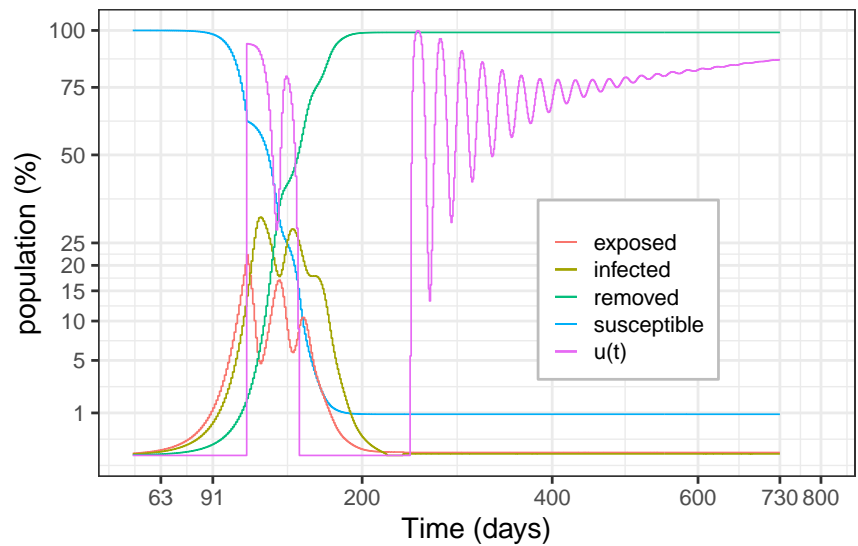}}
	\caption{\textbf{Epidemic evolution for Case F}}
	\label{fig:case_f}
\end{figure}

Case F illustrates the increasingly reduced flexibility that results from a belated mitigation. Observe in Figure \ref{fig:case_f} that the overall effect of the mitigation is limited. Although we are able to halve the peak of infections with respect to the unmitigated epidemic (Case A), with the total of simultaneous infections peaking at 31\% by day 121, the evolution of the epidemic still resembles the unmitigated case. High levels of control are demanded at the beginning of the mitigation period to contain the spike in infections. Still, as the removed population reaches about 50\%, the mitigation is dropped because the bulk of the population is either removed, infected or exposed. The disease then stabilises, but because 1\% of the population is still susceptible, enforced mitigation is necessary again to avoid another spike of the epidemic around day 250.

\begin{figure}[!htb]
	\centering
	\centerline{\includegraphics[width=\linewidth] {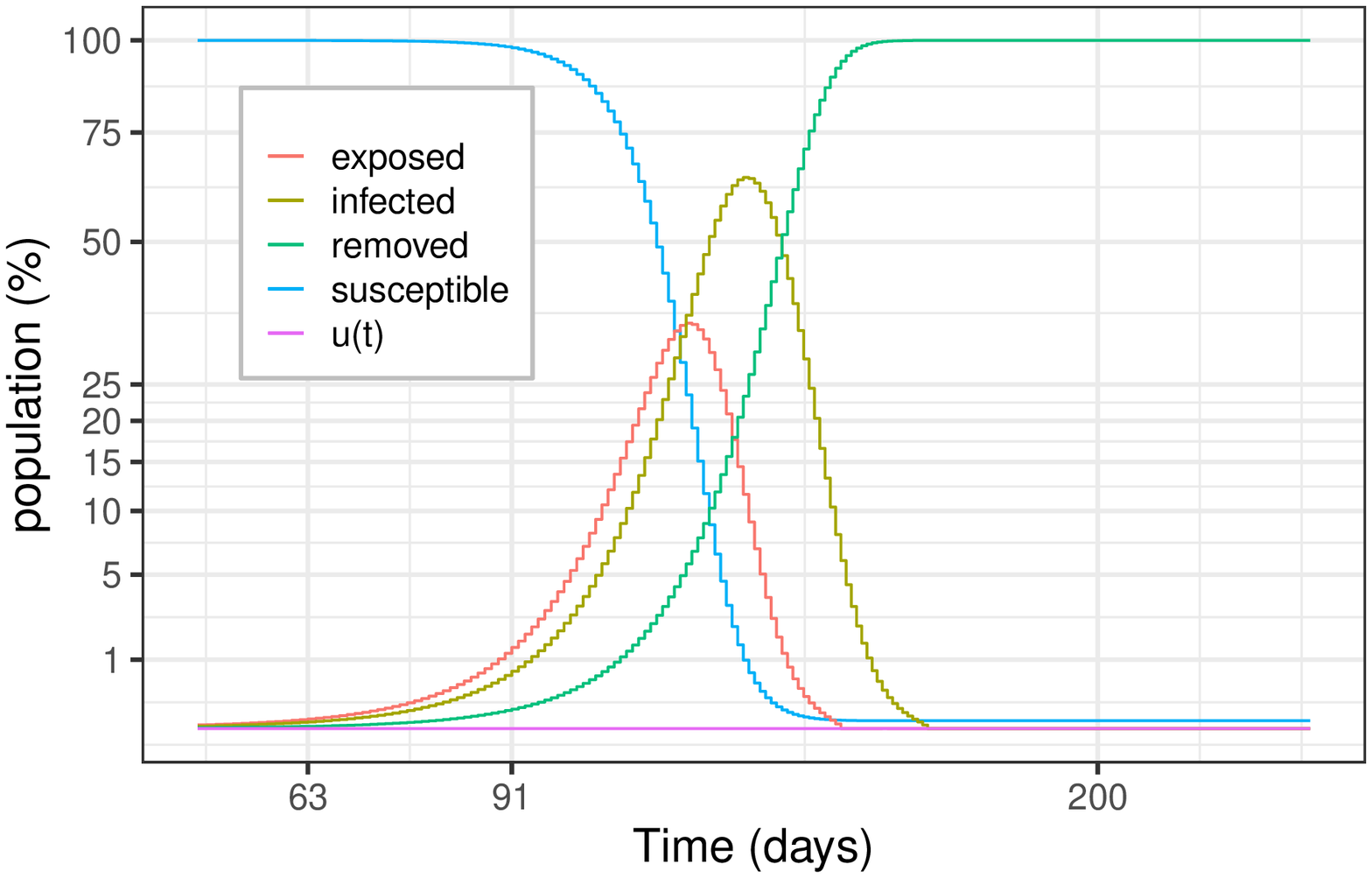}}
	\caption{\textbf{Case G}}
	\label{fig:case_g}
\end{figure}

Finally, Figure \ref{fig:case_g} presents the results for Case G, which demonstrate that delaying the mitigation by 126 days effectively renders it meaningless. Indeed, on day 126, the uncontrolled epidemic is so widespread that the infection levels are already dropping due to a small number of susceptible individuals remaining in the population. The infection reaches its peak on day 129 when the infection simultaneously afflicts approximately 64\% of the population.

\subsection{The effect of distinct occupation rates}

\begin{figure}[!htb]
	\centering
	\centerline{\includegraphics[width=\linewidth] {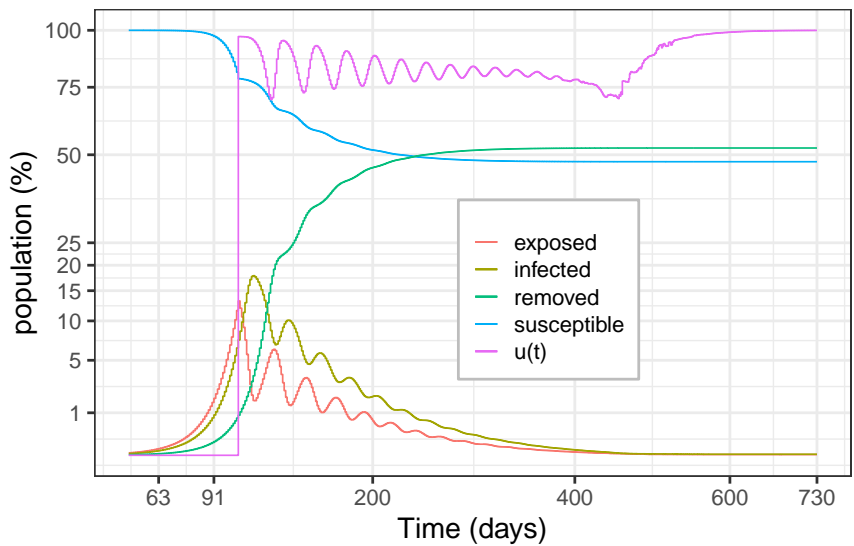}}
	\caption{\textbf{Epidemic evolution for Case H}}
	\label{fig:case_h}
\end{figure}

\begin{figure}[!htb]
	\centering
	\centerline{\includegraphics[width=\linewidth] {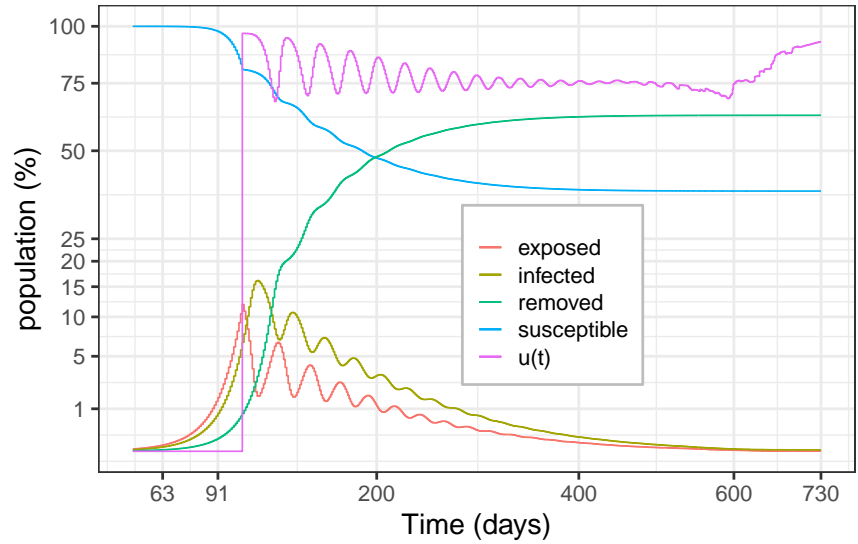}}
	\caption{\textbf{Epidemic evolution for Case I}}
	\label{fig:case_i}
\end{figure}

\begin{figure}[!htb]
	\centering
	\centerline{\includegraphics[width=\linewidth] {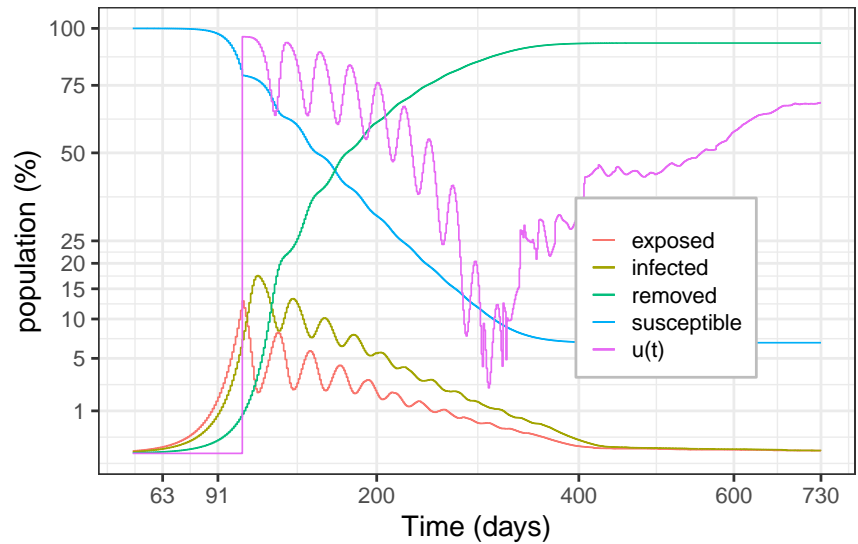}}
	\caption{\textbf{Epidemic evolution for Case J}}
	\label{fig:case_j}
\end{figure}

\begin{figure}[!htb]
	\centering
	\centerline{\includegraphics[width=\linewidth] {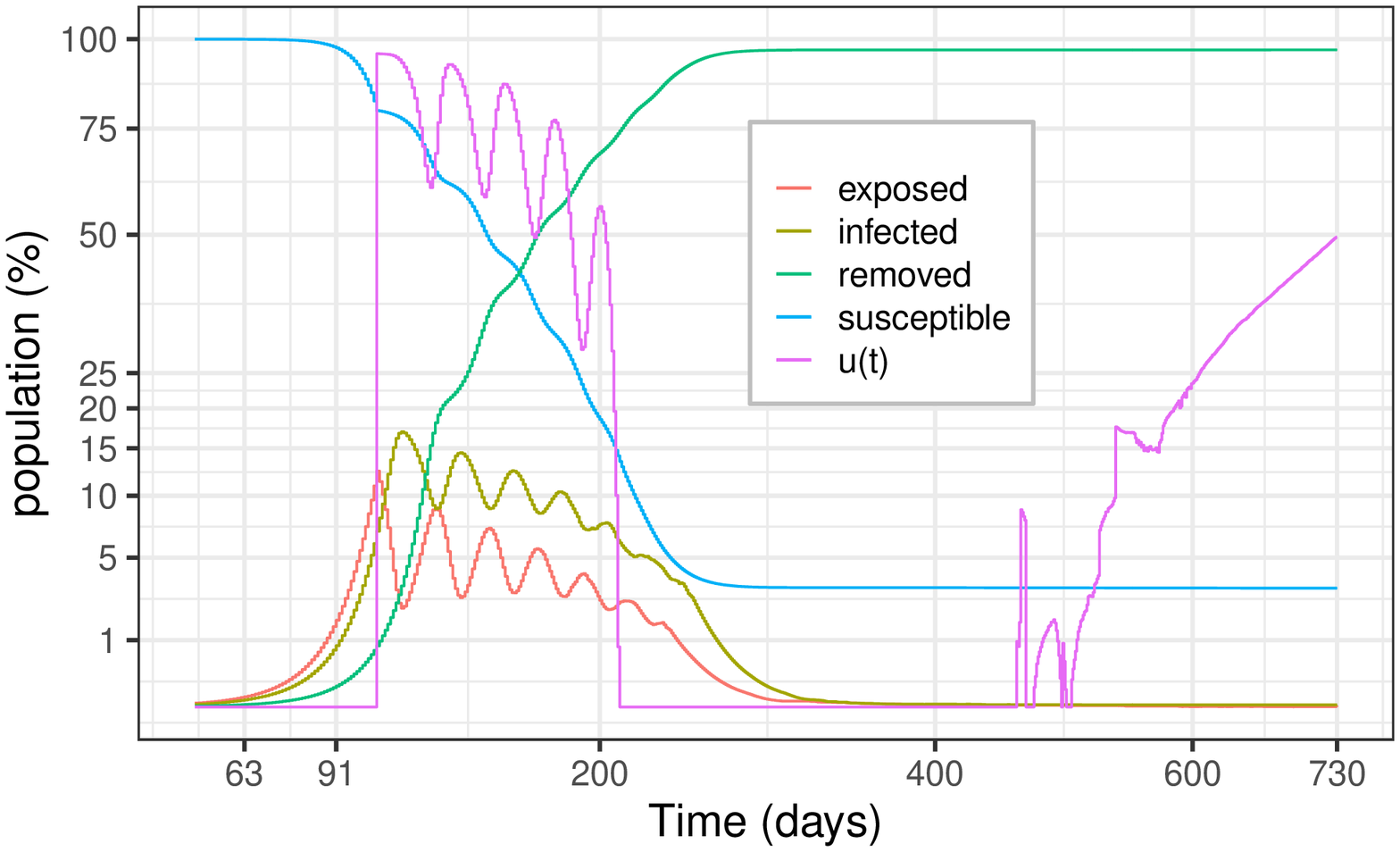}}
	\caption{\textbf{Epidemic evolution for Case K}}
	\label{fig:case_k}
\end{figure}

This Section investigates the effect of distinct target occupation rates $\bar \rho$ in the stabilising policies. To provide an interesting baseline, we start our mitigating actions on $\bar t = 105$, when the epidemic reaches a significant proportion of the population but can still be controlled as in Case E of Figure \ref{fig:case_e}. The set of experiments in Table \ref{tab:simcasestwo} assesses the effect of varying the target occupation level from 60\% to 90\% in regular intervals. To account for the stochastic fluctuations and prevent biases, Figures \ref{fig:case_h} to \ref{fig:case_k} show the median of 100 realisations of the stochastic system within a two-year interval.

\begin{center}
\begin{table}[htb]
\caption{\centering Second set of experiments. \label{tab:simcasestwo}}
\centering \scriptsize{
\begin{tabular}{|c|l|c|l|c|l|c|l|c|} \hline \hline
\textbf{Case} & H & I & J & K \\ \hline
\textbf{$\bar t$} & 105 & 105 & 105 & 105 \\ \hline
\textbf{$\bar \delta$} & 0.60 & 0.70 & 0.80 & 0.90 \\ \hline
\end{tabular}
}
\end{table}
\end{center}

Figure \ref{fig:case_h} presents the results for Case H. As one can observe, the number of infected people peaks around 18\% on day 114. As expected, the stronger mitigation has no effect in the early stages and is not able to prevent a similar peak as in the baseline (Case E). However, as time elapses we can notice that the increased mitigation produces a much more pronounced decrease in infections. In effect, infections drop to about 2\% by day 200 and 1\% on the $231^{th}$ day. The increased control is able to keep the removed population at about 52\% at the end of two years, whereas that level was 97.5\% in the baseline. It is noteworthy that the mitigation is kept at a high level after the system stabilises to avoid recrudescence.

The results for Case I are shown in Figure \ref{fig:case_i}. The increase in the target occupation rate $\bar \rho$ with respect to Case H has several  noticeable effects: a slight decrease in the mitigating actions; a slower rate of decrease of the infection levels and an increase of the removed population. The infection levels decrease to 3\% on day 200 and 72 days later reach 1\%. The reduced control levels also lead to an increase in the removed population, that climbs from 52\% in Case H to about 63\% in Case I.


The trend of increasing infection levels gathers speed in Case J, as the decrease in the required mitigating actions leads to about 93\% of removals within the two-year horizon. On day 200, the number of infections reaches 5\% of the population, and 117 days later it reaches 1\%. Coupled with the significant decrease in the susceptible population, which reduces the potential spread, the decrease in $\bar \rho$ produces a steep decrease in the control levels. These reach a minimum of 5\% as the susceptible population starts to stabilise. However, increased levels of mitigation are required later to avoid a second wave.

Finally, Figure \ref{fig:case_k} presents the results for Case K. As the target occupation level approaches that of the baseline (Case E), the overall behaviour becomes quite similar. We can notice a small decrease in the overall removals, that reach about 3.5\% as opposed to 2.5\% in the baseline. Despite dropping mitigating actions around day 210, the reduced size of the susceptible population causes the level of infection to fall sharply after reaching 5\% on day 237. After the system stabilises, late mitigation efforts are needed to keep the infection at bay just before the $500^{th}$ day.


\subsection{The effect of on-off lock-down policies}

This Section briefly evaluates the effect of \emph{on-off} lock-down policies. Proposed by \citet{Tarrataca2021}, these policies demand a full scale lock-down ($u(t)=1, \, \bar \rho=0$) when the number of infections surpass a prescribed upper limit; conversely, all measures are lifted ($u(t)=0, \, \bar \rho =\infty$) as soon as these numbers drop below a lower bound. We simulate the two policies whose upper and lower bounds appear in Table \ref{tab:simcasesthree}.

\begin{center}
\begin{table}[htb]
\caption{\centering Third set of experiments. \label{tab:simcasesthree}}
\centering \scriptsize{
\begin{tabular}{|c|l|c||} \hline \hline
\textbf{Case} & L & M \\ \hline
\textbf{$Start$} & 1,5\% & 2\% \\ \hline
\textbf{$Stop$} & 0.3\% & 1\% \\ \hline
\end{tabular}
}
\end{table}
\end{center}

\begin{figure}[!htb]
	\centering
	\centerline{\includegraphics[width=1.15\linewidth] {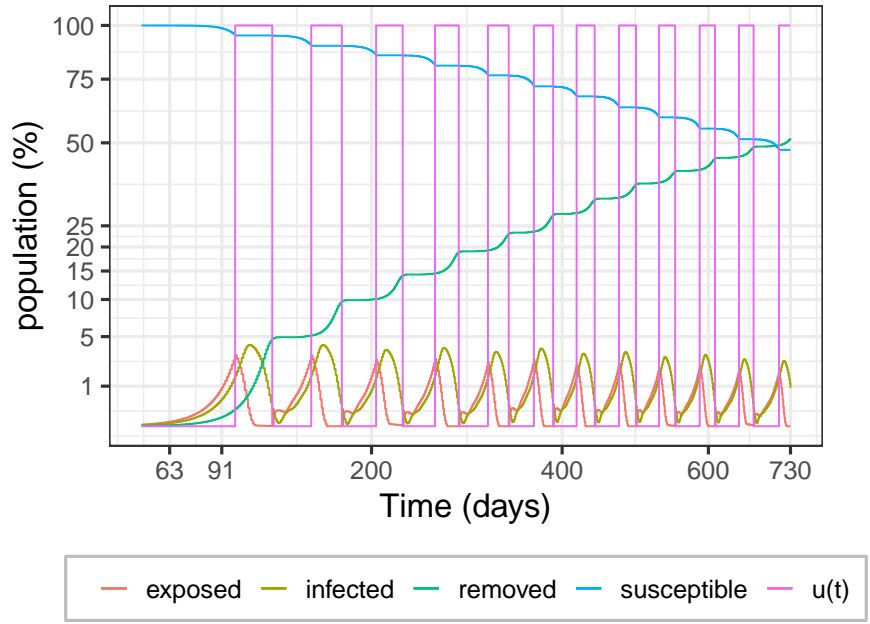}}
	\caption{\textbf{Epidemic evolution for Case L}}
	\label{fig:case_l}
\end{figure}

Figure \ref{fig:case_l} shows that Case L features 12 lock-downs in two years, the first starting on day 99. Infections recurrently exceed the upper bound in Table \ref{tab:simcasesthree} and peak around 4\% on day 108; the removed population amounts to 52\% at the end of the second year. Consistently with the results reported by \citet{Tarrataca2021}, the system alternates between rapid increases and decreases in infection and the lock-downs become shorter and more spaced over time.


\begin{figure}[!htb]
	\centering
	\centerline{\includegraphics[width=1.15\linewidth] {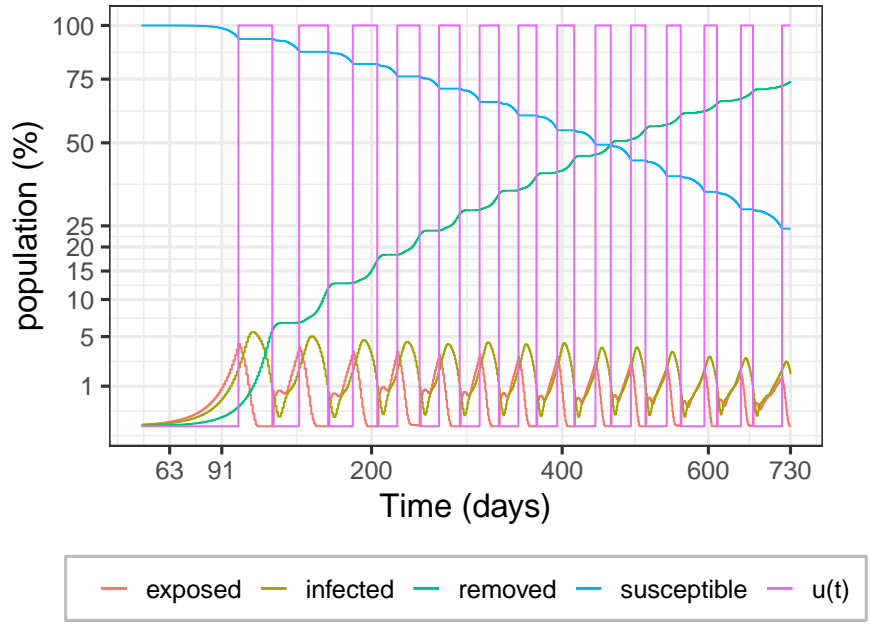}}
	\caption{\textbf{Epidemic evolution for Case M}}
	\label{fig:case_m}
\end{figure}

\begin{figure}[!htb]
	\centering
	\includegraphics[width=0.49\linewidth] {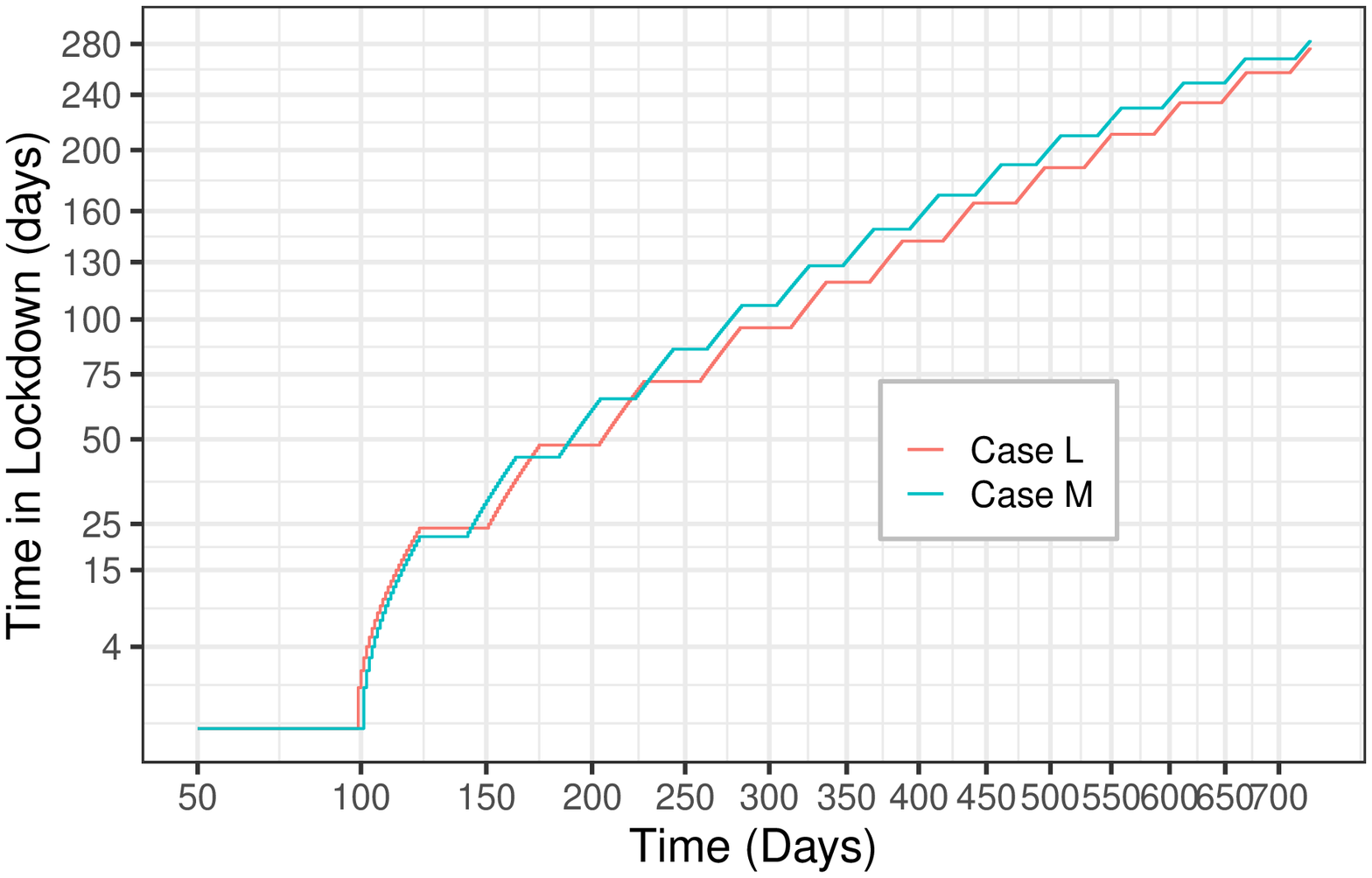}
	\includegraphics[width=0.49\linewidth] {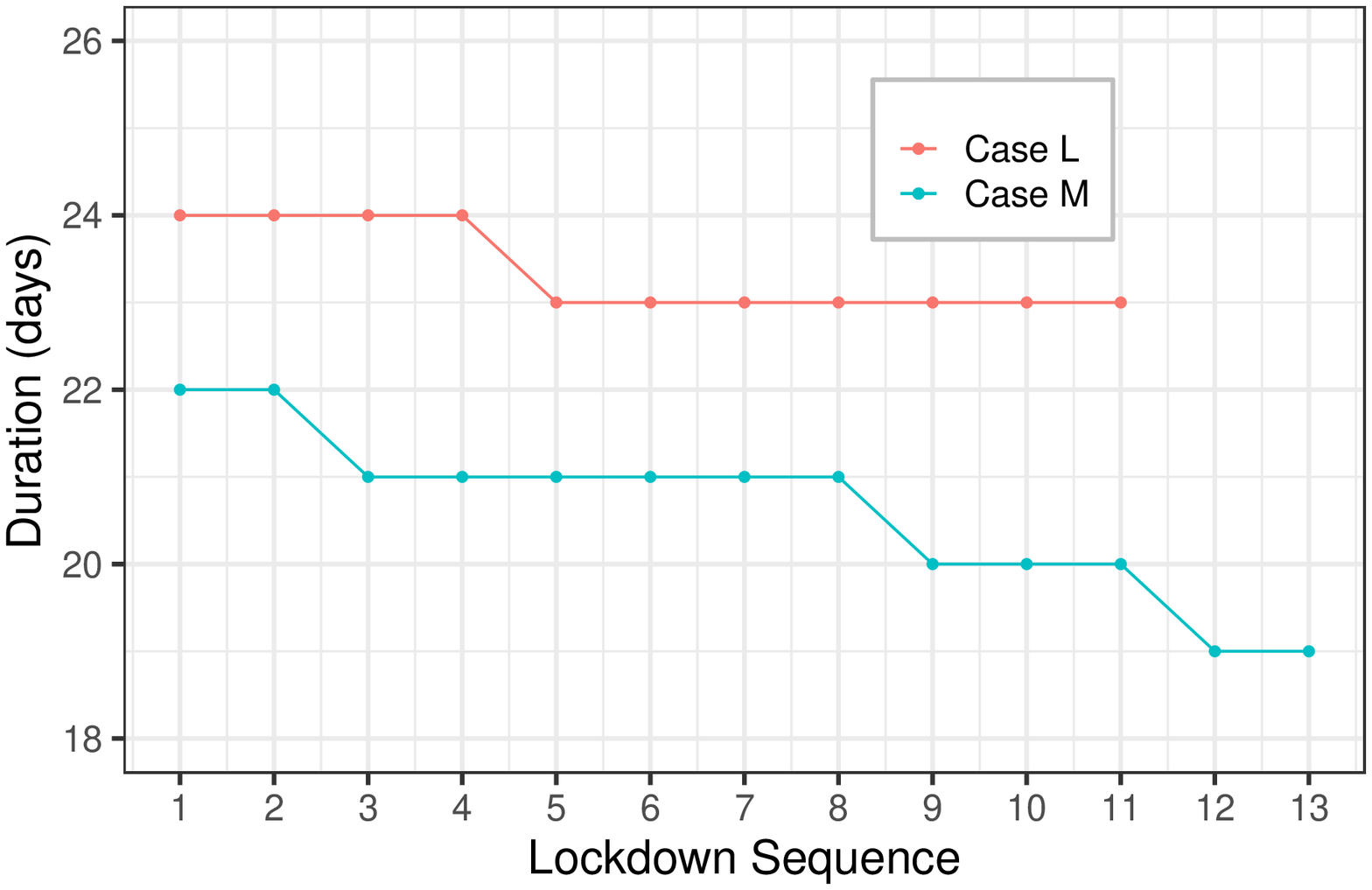}
	\caption{\textbf{Time in Lock-down in Cases L and M}}
	\label{fig:timesinlock}
\end{figure}

A similar behaviour is observed in Figure \ref{fig:case_m} for Case M. But because the upper and lower bounds are larger, the total number of removals increases from 52\% to 74\% in two years. In addition, the peak of infections observed on day 110 increases to 6\%. Although Case M has 2 more lock-downs in two years, the total time in lock-down is similar, as illustrated in Figure \ref{fig:timesinlock}, which shows the cumulative time in lock-down on the left hand side and the duration of individual lock-downs on the opposite half. The mean duration of a lock-down is about 23 days in case L and 21 days in case M, whereas the mean interval between lock-downs was 32 days for Case L and 27 days for Case M.

\section{Conclusions \label{sec:conc}}

Inspired by the unprecedented COVID-19 epidemic, this paper explored innovative stochastic modelling alternatives to describe the evolution of an epidemic by means of the classical SEIR framework. The proposed model innovates by simultaneously addressing general infection and latency periods, whilst the original methodology ensures that the resulting continuous-time Markov model remains tractable and useful to support decision making.

By drawing a parallel between stochastic stability and the traditional reproduction number, the paper introduced stabilising strategies that can curb an epidemic under very general conditions, provided that the mitigation is initiated in a timely manner. Beyond the academic contributions, we argue that the generality of the proposed approach renders it invaluable to support real-world decision making in the face of future epidemics. The methodology was validated using COVID-19 data from the literature. The results provide a panorama of insights into the pros and cons of distinct stabilising mitigation strategies over a two-year horizon.

As expected, the experiments illustrate that early intervention is vital to prevent the disease from affecting a large portion of the population. However, to effectively prevent the spread, we need very high levels of mitigation over the whole two-year horizon. The results also suggest that delayed mitigation leads to a dramatic increase in the overall number of infections. In effect, delays tend to produce persistent infection levels over time, even in the presence of mitigation, thereby increasing the spread even though the infection curves are effectively flattened. Beyond flattening the curves, this behaviour suggests that we also need to ensure, by acting swiftly, that their summit is tolerable from both societal and health care perspectives.

The proposed approach leaves many research avenues to be explored in future works. One obvious route is to pursue stochastic optimal control policies that somehow address the compromise between health care aspects, societal issues and the economic burden of mitigation strategies. The challenges involve finding a meaningful trade-off among the different elements that decision-makers need to consider, as well as proposing effective formulations that avoid the curse of dimensionality \citep{Powell2011,Powell2019} to ensure that the problem remains tractable. Another branch goes into developing filtering and analytical
approaches to estimate a system's parameters considering that the available information is delayed and biased, since dependent on local testing and reporting policies.

\section*{Acknowledgements}

This study was partly supported by the Brazilian Research Council—CNPq, under grants \#311075/2018-5, \#303352/2018-3 and 304801/2015-1, and by Coordenação de Aperfeiçoamento de Pessoal de Nível Superior—Brasil (CAPES) [Finance Code 001].



\scriptsize{

\begin{thebibliography}{30}
\expandafter\ifx\csname natexlab\endcsname\relax\def\natexlab#1{#1}\fi
\providecommand{\url}[1]{\texttt{#1}}
\providecommand{\href}[2]{#2}
\providecommand{\path}[1]{#1}
\providecommand{\DOIprefix}{doi:}
\providecommand{\ArXivprefix}{arXiv:}
\providecommand{\URLprefix}{URL: }
\providecommand{\Pubmedprefix}{pmid:}
\providecommand{\doi}[1]{\href{http://dx.doi.org/#1}{\path{#1}}}
\providecommand{\Pubmed}[1]{\href{pmid:#1}{\path{#1}}}
\providecommand{\bibinfo}[2]{#2}
\ifx\xfnm\relax \def\xfnm[#1]{\unskip,\space#1}\fi
\bibitem[{Allen(2008)}]{Allen2008}
\bibinfo{author}{Allen, L.J.S.}, \bibinfo{year}{2008}.
\newblock \bibinfo{title}{{An Introduction to Stochastic Epidemic Models}}, in:
  \bibinfo{editor}{Brauer, F.}, \bibinfo{editor}{{van den Driessche}, P.},
  \bibinfo{editor}{Wu, J.} (Eds.), \bibinfo{booktitle}{Mathematical
  Epidemiology}. \bibinfo{publisher}{Springer Berlin Heidelberg},
  \bibinfo{address}{Berlin, Heidelberg}, pp. \bibinfo{pages}{81--130}.
\newblock \DOIprefix\doi{10.1007/978-3-540-78911-6\_3}.
\bibitem[{Amador and Lopez-Herrero(2018)}]{Amador2018}
\bibinfo{author}{Amador, J.}, \bibinfo{author}{Lopez-Herrero, M.},
  \bibinfo{year}{2018}.
\newblock \bibinfo{title}{Cumulative and maximum epidemic sizes for a nonlinear
  {SEIR} stochastic model with limited resources}.
\newblock \bibinfo{journal}{Discrete \& Continuous Dynamical Systems - B}
  \bibinfo{volume}{23}, \bibinfo{pages}{3137}.
\newblock \DOIprefix\doi{10.3934/dcdsb.2017211}.
\bibitem[{Artalejo et~al.(2015)Artalejo, Economou and
  Lopez-Herrero}]{Artalejo2015}
\bibinfo{author}{Artalejo, J.R.}, \bibinfo{author}{Economou, A.},
  \bibinfo{author}{Lopez-Herrero, M.J.}, \bibinfo{year}{2015}.
\newblock \bibinfo{title}{The stochastic {SEIR }model before extinction:
  {Computational} approaches}.
\newblock \bibinfo{journal}{Applied Mathematics and Computation}
  \bibinfo{volume}{265}, \bibinfo{pages}{1026 -- 1043}.
\newblock \DOIprefix\doi{10.1016/j.amc.2015.05.141}.
\bibitem[{Backer et~al.(2020)Backer, Klinkenberg and Wallinga}]{Backer2020}
\bibinfo{author}{Backer, J.}, \bibinfo{author}{Klinkenberg, D.},
  \bibinfo{author}{Wallinga, J.}, \bibinfo{year}{2020}.
\newblock \bibinfo{title}{Incubation period of 2019 novel coronavirus
  {(2019-nCoV)} infections among travellers from {Wuhan, China}, 20–28
  january 2020}.
\newblock \bibinfo{journal}{Eurosurveillance} \bibinfo{volume}{25}.
\newblock \DOIprefix\doi{10.2807/1560-7917.ES.2020.25.5.2000062}.
\bibitem[{Barraza et~al.(2020)Barraza, Pena and Moreno}]{Barraza2020}
\bibinfo{author}{Barraza, N.R.}, \bibinfo{author}{Pena, G.},
  \bibinfo{author}{Moreno, V.}, \bibinfo{year}{2020}.
\newblock \bibinfo{title}{A non-homogeneous {Markov} early epidemic growth
  dynamics model. application to the {SARS-CoV-2} pandemic}.
\newblock \bibinfo{journal}{Chaos, Solitons \& Fractals} \bibinfo{volume}{139},
  \bibinfo{pages}{110297}.
\newblock \DOIprefix\doi{10.1016/j.chaos.2020.110297}.
\bibitem[{Br\'{e}maud(1999)}]{Bremaud1999}
\bibinfo{author}{Br\'{e}maud, P.}, \bibinfo{year}{1999}.
\newblock \bibinfo{title}{Gibbs fields, monte carlo simulation, and queues}.
\newblock \bibinfo{publisher}{Springer-Verlag}, \bibinfo{address}{New York}.
\bibitem[{Britton(2010)}]{Britton2010}
\bibinfo{author}{Britton, T.}, \bibinfo{year}{2010}.
\newblock \bibinfo{title}{Stochastic epidemic models: {A} survey}.
\newblock \bibinfo{journal}{Mathematical Biosciences} \bibinfo{volume}{225},
  \bibinfo{pages}{24 -- 35}.
\newblock \DOIprefix\doi{10.1016/j.mbs.2010.01.006}.
\bibitem[{Clancy(2014)}]{Clancy2014}
\bibinfo{author}{Clancy, D.}, \bibinfo{year}{2014}.
\newblock \bibinfo{title}{{SIR} epidemic models with general infectious period
  distribution}.
\newblock \bibinfo{journal}{Statistics \& Probability Letters}
  \bibinfo{volume}{85}, \bibinfo{pages}{1 -- 5}.
\newblock \DOIprefix\doi{10.1016/j.spl.2013.10.017}.
\bibitem[{Davis(1993)}]{Davis1993}
\bibinfo{author}{Davis, M.H.A.}, \bibinfo{year}{1993}.
\newblock \bibinfo{title}{{M}arkov models and optimization}.
\newblock \bibinfo{publisher}{Chapman and Hall}, \bibinfo{address}{London}.
\bibitem[{Dike et~al.(2016)Dike, Zainuddin and Dike}]{Dike2016}
\bibinfo{author}{Dike, C.O.}, \bibinfo{author}{Zainuddin, Z.M.},
  \bibinfo{author}{Dike, I.J.}, \bibinfo{year}{2016}.
\newblock \bibinfo{title}{{Queueing Technique for Ebola Virus Disease
  Transmission and Control Analysis}}.
\newblock \bibinfo{journal}{Indian Journal of Science and Technology}
  \bibinfo{volume}{9}.
\newblock \DOIprefix\doi{10.17485/ijst/2016/v9i46/107077}.
\bibitem[{Eick et~al.(1993)Eick, Massey and Whitt}]{Eick1993}
\bibinfo{author}{Eick, S.G.}, \bibinfo{author}{Massey, W.A.},
  \bibinfo{author}{Whitt, W.}, \bibinfo{year}{1993}.
\newblock \bibinfo{title}{{The Physics of the $M_t/G/\infty$ Queue}}.
\newblock \bibinfo{journal}{Operations Research} \bibinfo{volume}{41},
  \bibinfo{pages}{731--742}.
\newblock \DOIprefix\doi{10.1287/opre.41.4.731}.
\bibitem[{Ferguson et~al.(2020)Ferguson, Laydon, Nedjati~Gilani, Imai, Ainslie,
  Baguelin, Bhatia, Boonyasiri, Cucunuba~Perez, Cuomo-Dannenburg, Dighe,
  Dorigatti, Fu, Gaythorpe, Green, Hamlet, Hinsley, Okell, Van~Elsland,
  Thompson, Verity, Volz, Wang, Wang, Walker, Winskill, Whittaker, Donnelly,
  Riley and Ghani}]{ferguson2020}
\bibinfo{author}{Ferguson, N.}, \bibinfo{author}{Laydon, D.},
  \bibinfo{author}{Nedjati~Gilani, G.}, \bibinfo{author}{Imai, N.},
  \bibinfo{author}{Ainslie, K.}, \bibinfo{author}{Baguelin, M.},
  \bibinfo{author}{Bhatia, S.}, \bibinfo{author}{Boonyasiri, A.},
  \bibinfo{author}{Cucunuba~Perez, Z.}, \bibinfo{author}{Cuomo-Dannenburg, G.},
  \bibinfo{author}{Dighe, A.}, \bibinfo{author}{Dorigatti, I.},
  \bibinfo{author}{Fu, H.}, \bibinfo{author}{Gaythorpe, K.},
  \bibinfo{author}{Green, W.}, \bibinfo{author}{Hamlet, A.},
  \bibinfo{author}{Hinsley, W.}, \bibinfo{author}{Okell, L.},
  \bibinfo{author}{Van~Elsland, S.}, \bibinfo{author}{Thompson, H.},
  \bibinfo{author}{Verity, R.}, \bibinfo{author}{Volz, E.},
  \bibinfo{author}{Wang, H.}, \bibinfo{author}{Wang, Y.},
  \bibinfo{author}{Walker, P.}, \bibinfo{author}{Winskill, P.},
  \bibinfo{author}{Whittaker, C.}, \bibinfo{author}{Donnelly, C.},
  \bibinfo{author}{Riley, S.}, \bibinfo{author}{Ghani, A.},
  \bibinfo{year}{2020}.
\newblock \bibinfo{title}{Report 9: Impact of non-pharmaceutical interventions
  {(NPIs)} to reduce {COVID-19} mortality and healthcare demand}.
\newblock \bibinfo{type}{Technical Report}. Imperial College London.
\newblock \DOIprefix\doi{10.25561/77482}.
\bibitem[{Gómez-Corral and López-García(2017)}]{Corral2017}
\bibinfo{author}{Gómez-Corral, A.}, \bibinfo{author}{López-García, M.},
  \bibinfo{year}{2017}.
\newblock \bibinfo{title}{On {SIR} epidemic models with generally distributed
  infectious periods: Number of secondary cases and probability of infection}.
\newblock \bibinfo{journal}{International Journal of Biomathematics}
  \bibinfo{volume}{10}, \bibinfo{pages}{1750024}.
\newblock \DOIprefix\doi{10.1142/S1793524517500243}.
\bibitem[{Hethcote(2000)}]{Hethcote2000}
\bibinfo{author}{Hethcote, H.W.}, \bibinfo{year}{2000}.
\newblock \bibinfo{title}{{The Mathematics of Infectious Diseases}}.
\newblock \bibinfo{journal}{SIAM Review} \bibinfo{volume}{42},
  \bibinfo{pages}{599--653}.
\newblock \DOIprefix\doi{10.1137/S0036144500371907}.
\bibitem[{Kantner and Koprucki(2020)}]{Kantner2020}
\bibinfo{author}{Kantner, M.}, \bibinfo{author}{Koprucki, T.},
  \bibinfo{year}{2020}.
\newblock \bibinfo{title}{Beyond just ``flattening the curve'': Optimal control
  of epidemics with purely non-pharmaceutical interventions}.
\newblock \bibinfo{journal}{Journal of Mathematics in Industry}
  \bibinfo{volume}{10}, \bibinfo{pages}{23}.
\newblock \DOIprefix\doi{10.1186/s13362-020-00091-3}.
\bibitem[{Kermack et~al.(1927)Kermack, McKendrick and Walker}]{Kermack1927}
\bibinfo{author}{Kermack, W.O.}, \bibinfo{author}{McKendrick, A.G.},
  \bibinfo{author}{Walker, G.T.}, \bibinfo{year}{1927}.
\newblock \bibinfo{title}{A contribution to the mathematical theory of
  epidemics}.
\newblock \bibinfo{journal}{Proceedings of the Royal Society of London. Series
  A, Containing Papers of a Mathematical and Physical Character}
  \bibinfo{volume}{115}, \bibinfo{pages}{700--721}.
\newblock \DOIprefix\doi{10.1098/rspa.1927.0118}.
\bibitem[{Lef{\`e}vre and Simon(2020)}]{Lefevre2020}
\bibinfo{author}{Lef{\`e}vre, C.}, \bibinfo{author}{Simon, M.},
  \bibinfo{year}{2020}.
\newblock \bibinfo{title}{{SIR-Type Epidemic Models as Block-Structured Markov
  Processes}}.
\newblock \bibinfo{journal}{Methodology and Computing in Applied Probability}
  \bibinfo{volume}{22}, \bibinfo{pages}{433--453}.
\newblock \DOIprefix\doi{10.1007/s11009-019-09710-y}.
\bibitem[{Lopez-Herrero(2017)}]{Lopez2017}
\bibinfo{author}{Lopez-Herrero, M.}, \bibinfo{year}{2017}.
\newblock \bibinfo{title}{{Epidemic Transmission on SEIR Stochastic Models with
  Nonlinear Incidence Rate}}.
\newblock \bibinfo{journal}{Mathematical Methods in the Applied Sciences}
  \bibinfo{volume}{40}, \bibinfo{pages}{2532--2541}.
\newblock \DOIprefix\doi{10.1002/mma.4179}.
\bibitem[{López-García(2016)}]{Lopez2016}
\bibinfo{author}{López-García, M.}, \bibinfo{year}{2016}.
\newblock \bibinfo{title}{Stochastic descriptors in an {SIR} epidemic model for
  heterogeneous individuals in small networks}.
\newblock \bibinfo{journal}{Mathematical Biosciences} \bibinfo{volume}{271},
  \bibinfo{pages}{42 -- 61}.
\newblock \DOIprefix\doi{10.1016/j.mbs.2015.10.010}.
\bibitem[{Meyn and Tweedie(1993)}]{meyn93}
\bibinfo{author}{Meyn, S.P.}, \bibinfo{author}{Tweedie, R.L.},
  \bibinfo{year}{1993}.
\newblock \bibinfo{title}{{Markov Chains and Stochastic Stability}}.
\newblock \bibinfo{publisher}{Springer-Verlag}, \bibinfo{address}{New York}.
\bibitem[{Perkins and Espa{\~n}a(2020)}]{Perkins2020}
\bibinfo{author}{Perkins, T.A.}, \bibinfo{author}{Espa{\~n}a, G.},
  \bibinfo{year}{2020}.
\newblock \bibinfo{title}{{Optimal Control of the COVID-19 Pandemic with
  Non-pharmaceutical Interventions}}.
\newblock \bibinfo{journal}{Bulletin of Mathematical Biology}
  \bibinfo{volume}{82}, \bibinfo{pages}{118}.
\newblock \DOIprefix\doi{10.1007/s11538-020-00795-y}.
\bibitem[{Powell(2011)}]{Powell2011}
\bibinfo{author}{Powell, W.}, \bibinfo{year}{2011}.
\newblock \bibinfo{title}{{Approximate Dynamic Programming Solving the Curses
  of Dimensionality}}.
\newblock \bibinfo{publisher}{John Wiley {\&} Sons, Inc.},
  \bibinfo{address}{New Jersey, USA}.
\bibitem[{Powell(2019)}]{Powell2019}
\bibinfo{author}{Powell, W.}, \bibinfo{year}{2019}.
\newblock \bibinfo{title}{{A Unified Framework for Stochastic Optimization}}.
\newblock \bibinfo{journal}{European Journal of Operational Research}
  \bibinfo{volume}{275}, \bibinfo{pages}{795--821}.
\newblock \DOIprefix\doi{10.1016/J.EJOR.2018.07.014}.
\bibitem[{Ross(1916)}]{Ross1916}
\bibinfo{author}{Ross, R.}, \bibinfo{year}{1916}.
\newblock \bibinfo{title}{An application of the theory of probabilities to the
  study of a priori pathometry-part {I}}.
\newblock \bibinfo{journal}{Proceedings of the Royal Society of London. Series
  A, Containing Papers of a Mathematical and Physical Character}
  \bibinfo{volume}{92}, \bibinfo{pages}{204--230}.
\newblock \DOIprefix\doi{10.1098/rspa.1916.0007}.
\bibitem[{Shortle et~al.(2018)Shortle, Thompson, Gross and
  Harris}]{Shortle2018}
\bibinfo{author}{Shortle, J.}, \bibinfo{author}{Thompson, J.},
  \bibinfo{author}{Gross, D.}, \bibinfo{author}{Harris, C.},
  \bibinfo{year}{2018}.
\newblock \bibinfo{title}{{Fundamentals of Queueing Theory}}.
\newblock Wiley Series in Probability and Statistics. \bibinfo{edition}{5} ed.,
  \bibinfo{publisher}{Wiley}, \bibinfo{address}{New York}.
\newblock \DOIprefix\doi{10.1002/9781119453765}.
\bibitem[{Tarrataca et~al.(2021)Tarrataca, Dias, Haddad and
  Arruda}]{Tarrataca2021}
\bibinfo{author}{Tarrataca, L.}, \bibinfo{author}{Dias, C.M.},
  \bibinfo{author}{Haddad, D.}, \bibinfo{author}{Arruda, E.F.},
  \bibinfo{year}{2021}.
\newblock \bibinfo{title}{{Flattening the curves: on-off lock-down strategies
  for COVID-19 with an application to Brazil}}.
\newblock \bibinfo{journal}{Journal of Mathematics in Industry}
  \bibinfo{volume}{11}, \bibinfo{pages}{2}.
\newblock \DOIprefix\doi{10.1186/s13362-020-00098-w}.
\bibitem[{Trapman and Bootsma(2009)}]{Trapman2009}
\bibinfo{author}{Trapman, P.}, \bibinfo{author}{Bootsma, M.C.J.},
  \bibinfo{year}{2009}.
\newblock \bibinfo{title}{A useful relationship between epidemiology and
  queueing theory: {The} distribution of the number of infectives at the moment
  of the first detection}.
\newblock \bibinfo{journal}{Mathematical Biosciences} \bibinfo{volume}{219},
  \bibinfo{pages}{15 -- 22}.
\newblock \DOIprefix\doi{10.1016/j.mbs.2009.02.001}.
\bibitem[{{van den Driessche} and Watmough(2002)}]{Driessche2002}
\bibinfo{author}{{van den Driessche}, P.}, \bibinfo{author}{Watmough, J.},
  \bibinfo{year}{2002}.
\newblock \bibinfo{title}{Reproduction numbers and sub-threshold endemic
  equilibria for compartmental models of disease transmission}.
\newblock \bibinfo{journal}{Mathematical Biosciences} \bibinfo{volume}{180},
  \bibinfo{pages}{29 -- 48}.
\newblock \DOIprefix\doi{https://doi.org/10.1016/S0025-5564(02)00108-6}.
\bibitem[{Verity et~al.(2020)Verity, Okell, Dorigatti, Winskill, Whittaker,
  Imai, Cuomo-Dannenburg, Thompson, Walker, Fu, Dighe, Griffin, Baguelin,
  Bhatia, Boonyasiri, Cori, Cucunubá, FitzJohn, Gaythorpe, Green, Hamlet,
  Hinsley, Laydon, Nedjati-Gilani, Riley, {van Elsland}, Volz, Wang, Wang, Xi,
  Donnelly, Ghani and Ferguson}]{Verity2020}
\bibinfo{author}{Verity, R.}, \bibinfo{author}{Okell, L.C.},
  \bibinfo{author}{Dorigatti, I.}, \bibinfo{author}{Winskill, P.},
  \bibinfo{author}{Whittaker, C.}, \bibinfo{author}{Imai, N.},
  \bibinfo{author}{Cuomo-Dannenburg, G.}, \bibinfo{author}{Thompson, H.},
  \bibinfo{author}{Walker, P.G.T.}, \bibinfo{author}{Fu, H.},
  \bibinfo{author}{Dighe, A.}, \bibinfo{author}{Griffin, J.T.},
  \bibinfo{author}{Baguelin, M.}, \bibinfo{author}{Bhatia, S.},
  \bibinfo{author}{Boonyasiri, A.}, \bibinfo{author}{Cori, A.},
  \bibinfo{author}{Cucunubá, Z.}, \bibinfo{author}{FitzJohn, R.},
  \bibinfo{author}{Gaythorpe, K.}, \bibinfo{author}{Green, W.},
  \bibinfo{author}{Hamlet, A.}, \bibinfo{author}{Hinsley, W.},
  \bibinfo{author}{Laydon, D.}, \bibinfo{author}{Nedjati-Gilani, G.},
  \bibinfo{author}{Riley, S.}, \bibinfo{author}{{van Elsland}, S.},
  \bibinfo{author}{Volz, E.}, \bibinfo{author}{Wang, H.},
  \bibinfo{author}{Wang, Y.}, \bibinfo{author}{Xi, X.},
  \bibinfo{author}{Donnelly, C.A.}, \bibinfo{author}{Ghani, A.C.},
  \bibinfo{author}{Ferguson, N.M.}, \bibinfo{year}{2020}.
\newblock \bibinfo{title}{Estimates of the severity of coronavirus disease
  2019: a model-based analysis}.
\newblock \bibinfo{journal}{The Lancet Infectious Diseases}
  \DOIprefix\doi{10.1016/S1473-3099(20)30243-7}.
\bibitem[{İşlier et~al.(2020)İşlier, Güllü and Hörmann}]{Islier2020}
\bibinfo{author}{İşlier, Z.G.}, \bibinfo{author}{Güllü, R.},
  \bibinfo{author}{Hörmann, W.}, \bibinfo{year}{2020}.
\newblock \bibinfo{title}{An exact and implementable computation of the final
  outbreak size distribution under {Erlang} distributed infectious period}.
\newblock \bibinfo{journal}{Mathematical Biosciences} \bibinfo{volume}{325},
  \bibinfo{pages}{108363}.
\newblock \DOIprefix\doi{10.1016/j.mbs.2020.108363}.

\end{thebibliography}

}







\end{document}